\documentclass[prd,reprint,showpacs,superscriptaddress,bibnotes,floatfix]{revtex4-1}
\usepackage{natbib}
\usepackage{graphics}
\usepackage{multirow}
\usepackage{amsbsy}
\usepackage{mathrsfs}
\usepackage{footmisc}
\usepackage{hyperref}
\hypersetup{
  colorlinks = true,
  urlcolor = black,
  linkcolor=black,
  citecolor=black
}

\def\Ad{A_\mathrm{d}}
\def\Adf{A_\mathrm{d,353}}
\def\As{A_\mathrm{sync}}
\def\Asf{A_\mathrm{sync,23}}

\def\Bd{\beta_\mathrm{d}}
\def\Bs{\beta_\mathrm{s}}

\def\ad{\alpha_\mathrm{d}}
\def\as{\alpha_\mathrm{s}}

\def\bicep{BICEP}

\def\bicepone{{\sc BICEP1}}
\def\biceptwo{{\sc BICEP2}}
\def\bicepthree{{\sc BICEP3}}

\def\keck{{\it Keck}}
\def\keckarray{{\it Keck Array}}

\def\planck{{\it Planck}} 

\def\wmap{WMAP}

\def\deg{^\circ}
\def\emode{$E$-mode}
\def\bmode{$B$-mode}

\def\clstar{\ell \left( \ell + 1 \right) C_l / 2 \pi}
\def\lcdm{$\Lambda$CDM}

\def\aap{Astron.\ Astrophys.}

\def\apj{Astrophys.\ J.}
\def\apjl{Astrophys.\ J.\ Lett.}
\def\apjs{Astrophys.\ J.\ Suppl.\ Ser.}

\def\jcap{J.\ Cosmol.\ Astropart.\ Phys.}

\def\nat{Nature}

\def\prd{Phys.\ Rev.\ D}

\begin{document}

\title{\biceptwo\ / \keckarray\ VI:
Improved Constraints On Cosmology and Foregrounds When Adding 95\,GHz Data From \keckarray}

\author{\keckarray\ and \biceptwo\ Collaborations: P.~A.~R.~Ade}
\affiliation{School of Physics and Astronomy, Cardiff University, Cardiff, CF24 3AA, United Kingdom}
\author{Z.~Ahmed}
\affiliation{Kavli Institute for Particle Astrophysics and Cosmology, SLAC National Accelerator Laboratory, 2575 Sand Hill Rd, Menlo Park, California 94025, USA}
\affiliation{Department of Physics, Stanford University, Stanford, California 94305, USA}
\author{R.~W.~Aikin}
\affiliation{Department of Physics, California Institute of Technology, Pasadena, California 91125, USA}
\author{K.~D.~Alexander}
\affiliation{Harvard-Smithsonian Center for Astrophysics, 60 Garden Street MS 42, Cambridge, Massachusetts 02138, USA}
\author{D.~Barkats}
\affiliation{Harvard-Smithsonian Center for Astrophysics, 60 Garden Street MS 42, Cambridge, Massachusetts 02138, USA}
\author{S.~J.~Benton}
\affiliation{Department of Physics, University of Toronto, Toronto, Ontario, M5S 1A7, Canada}
\author{C.~A.~Bischoff}
\affiliation{Harvard-Smithsonian Center for Astrophysics, 60 Garden Street MS 42, Cambridge, Massachusetts 02138, USA}
\author{J.~J.~Bock}
\affiliation{Department of Physics, California Institute of Technology, Pasadena, California 91125, USA}
\affiliation{Jet Propulsion Laboratory, Pasadena, California 91109, USA}
\author{R.~Bowens-Rubin}
\affiliation{Harvard-Smithsonian Center for Astrophysics, 60 Garden Street MS 42, Cambridge, Massachusetts 02138, USA}
\author{J.~A.~Brevik}
\affiliation{Department of Physics, California Institute of Technology, Pasadena, California 91125, USA}
\author{I.~Buder}
\affiliation{Harvard-Smithsonian Center for Astrophysics, 60 Garden Street MS 42, Cambridge, Massachusetts 02138, USA}
\author{E.~Bullock}
\affiliation{Minnesota Institute for Astrophysics, University of Minnesota, Minneapolis, Minnesota 55455, USA}
\author{V.~Buza}
\affiliation{Harvard-Smithsonian Center for Astrophysics, 60 Garden Street MS 42, Cambridge, Massachusetts 02138, USA}
\affiliation{Department of Physics, Harvard University, Cambridge, MA 02138, USA}
\author{J.~Connors}
\affiliation{Harvard-Smithsonian Center for Astrophysics, 60 Garden Street MS 42, Cambridge, Massachusetts 02138, USA}
\author{B.~P.~Crill}
\affiliation{Jet Propulsion Laboratory, Pasadena, California 91109, USA}
\author{L.~Duband}
\affiliation{Service des Basses Temp\'{e}ratures, Commissariat \`{a} l'Energie Atomique, 38054 Grenoble, France}
\author{C.~Dvorkin}
\affiliation{Department of Physics, Harvard University, Cambridge, MA 02138, USA}
\author{J.~P.~Filippini}
\affiliation{Department of Physics, California Institute of Technology, Pasadena, California 91125, USA}
\affiliation{Department of Physics, University of Illinois at Urbana-Champaign, Urbana, Illinois 61801, USA}
\author{S.~Fliescher}
\affiliation{School of Physics and Astronomy, University of Minnesota, Minneapolis, Minnesota 55455, USA}
\author{J.~Grayson}
\affiliation{Department of Physics, Stanford University, Stanford, California 94305, USA}
\author{M.~Halpern}
\affiliation{Department of Physics and Astronomy, University of British Columbia, Vancouver, British Columbia, V6T 1Z1, Canada}
\author{S.~Harrison}
\affiliation{Harvard-Smithsonian Center for Astrophysics, 60 Garden Street MS 42, Cambridge, Massachusetts 02138, USA}
\author{G.~C.~Hilton}
\affiliation{National Institute of Standards and Technology, Boulder, Colorado 80305, USA}
\author{H.~Hui}
\affiliation{Department of Physics, California Institute of Technology, Pasadena, California 91125, USA}
\author{K.~D.~Irwin}
\affiliation{Department of Physics, Stanford University, Stanford, California 94305, USA}
\affiliation{Kavli Institute for Particle Astrophysics and Cosmology, SLAC National Accelerator Laboratory, 2575 Sand Hill Rd, Menlo Park, California 94025, USA}
\affiliation{National Institute of Standards and Technology, Boulder, Colorado 80305, USA}
\author{K.~S.~Karkare}
\affiliation{Harvard-Smithsonian Center for Astrophysics, 60 Garden Street MS 42, Cambridge, Massachusetts 02138, USA}
\author{E.~Karpel}
\affiliation{Department of Physics, Stanford University, Stanford, California 94305, USA}
\author{J.~P.~Kaufman}
\affiliation{Department of Physics, University of California at San Diego, La Jolla, California 92093, USA}
\author{B.~G.~Keating}
\affiliation{Department of Physics, University of California at San Diego, La Jolla, California 92093, USA}
\author{S.~Kefeli}
\affiliation{Department of Physics, California Institute of Technology, Pasadena, California 91125, USA}
\author{S.~A.~Kernasovskiy}
\affiliation{Department of Physics, Stanford University, Stanford, California 94305, USA}
\author{J.~M.~Kovac}
\email{jmkovac@cfa.harvard.edu}
\affiliation{Harvard-Smithsonian Center for Astrophysics, 60 Garden Street MS 42, Cambridge, Massachusetts 02138, USA}
\affiliation{Department of Physics, Harvard University, Cambridge, MA 02138, USA}
\author{C.~L.~Kuo}
\affiliation{Department of Physics, Stanford University, Stanford, California 94305, USA}
\affiliation{Kavli Institute for Particle Astrophysics and Cosmology, SLAC National Accelerator Laboratory, 2575 Sand Hill Rd, Menlo Park, California 94025, USA}
\author{E.~M.~Leitch}
\affiliation{Kavli Institute for Cosmological Physics, University of Chicago, Chicago, IL 60637, USA}
\author{M.~Lueker}
\affiliation{Department of Physics, California Institute of Technology, Pasadena, California 91125, USA}
\author{K.~G.~Megerian}
\affiliation{Jet Propulsion Laboratory, Pasadena, California 91109, USA}
\author{C.~B.~Netterfield}
\affiliation{Department of Physics, University of Toronto, Toronto, Ontario, M5S 1A7, Canada}
\affiliation{Canadian Institute for Advanced Research, Toronto, Ontario, M5G 1Z8, Canada}
\author{H.~T.~Nguyen}
\affiliation{Jet Propulsion Laboratory, Pasadena, California 91109, USA}
\author{R.~O'Brient}
\affiliation{Department of Physics, California Institute of Technology, Pasadena, California 91125, USA}
\affiliation{Jet Propulsion Laboratory, Pasadena, California 91109, USA}
\author{R.~W.~Ogburn~IV}
\affiliation{Department of Physics, Stanford University, Stanford, California 94305, USA}
\affiliation{Kavli Institute for Particle Astrophysics and Cosmology, SLAC National Accelerator Laboratory, 2575 Sand Hill Rd, Menlo Park, California 94025, USA}
\author{A.~Orlando}
\affiliation{Department of Physics, California Institute of Technology, Pasadena, California 91125, USA}
\affiliation{Department of Physics, University of California at San Diego, La Jolla, California 92093, USA}
\author{C.~Pryke}
\email{pryke@physics.umn.edu}
\affiliation{School of Physics and Astronomy, University of Minnesota, Minneapolis, Minnesota 55455, USA}
\affiliation{Minnesota Institute for Astrophysics, University of Minnesota, Minneapolis, Minnesota 55455, USA}
\author{S.~Richter}
\affiliation{Harvard-Smithsonian Center for Astrophysics, 60 Garden Street MS 42, Cambridge, Massachusetts 02138, USA}
\author{R.~Schwarz}
\affiliation{School of Physics and Astronomy, University of Minnesota, Minneapolis, Minnesota 55455, USA}
\author{C.~D.~Sheehy}
\affiliation{School of Physics and Astronomy, University of Minnesota, Minneapolis, Minnesota 55455, USA}
\affiliation{Kavli Institute for Cosmological Physics, University of Chicago, Chicago, IL 60637, USA}
\author{Z.~K.~Staniszewski}
\affiliation{Department of Physics, California Institute of Technology, Pasadena, California 91125, USA}
\affiliation{Jet Propulsion Laboratory, Pasadena, California 91109, USA}
\author{B.~Steinbach}
\affiliation{Department of Physics, California Institute of Technology, Pasadena, California 91125, USA}
\author{R.~V.~Sudiwala}
\affiliation{School of Physics and Astronomy, Cardiff University, Cardiff, CF24 3AA, United Kingdom}
\author{G.~P.~Teply}
\affiliation{Department of Physics, California Institute of Technology, Pasadena, California 91125, USA}
\affiliation{Department of Physics, University of California at San Diego, La Jolla, California 92093, USA}
\author{K.~L.~Thompson}
\affiliation{Department of Physics, Stanford University, Stanford, California 94305, USA}
\affiliation{Kavli Institute for Particle Astrophysics and Cosmology, SLAC National Accelerator Laboratory, 2575 Sand Hill Rd, Menlo Park, California 94025, USA}
\author{J.~E.~Tolan}
\affiliation{Department of Physics, Stanford University, Stanford, California 94305, USA}
\author{C.~Tucker}
\affiliation{School of Physics and Astronomy, Cardiff University, Cardiff, CF24 3AA, United Kingdom}
\author{A.~D.~Turner}
\affiliation{Jet Propulsion Laboratory, Pasadena, California 91109, USA}
\author{A.~G.~Vieregg}
\affiliation{Harvard-Smithsonian Center for Astrophysics, 60 Garden Street MS 42, Cambridge, Massachusetts 02138, USA}
\affiliation{Department of Physics, Enrico Fermi Institute, University of Chicago, Chicago, IL 60637, USA}
\affiliation{Kavli Institute for Cosmological Physics, University of Chicago, Chicago, IL 60637, USA}
\author{A.~C.~Weber}
\affiliation{Jet Propulsion Laboratory, Pasadena, California 91109, USA}
\author{D.~V.~Wiebe}
\affiliation{Department of Physics and Astronomy, University of British Columbia, Vancouver, British Columbia, V6T 1Z1, Canada}
\author{J.~Willmert}
\affiliation{School of Physics and Astronomy, University of Minnesota, Minneapolis, Minnesota 55455, USA}
\author{C.~L.~Wong}
\affiliation{Harvard-Smithsonian Center for Astrophysics, 60 Garden Street MS 42, Cambridge, Massachusetts 02138, USA}
\affiliation{Department of Physics, Harvard University, Cambridge, MA 02138, USA}
\author{W.~L.~K.~Wu}
\affiliation{Department of Physics, Stanford University, Stanford, California 94305, USA}
\author{K.~W.~Yoon}
\affiliation{Department of Physics, Stanford University, Stanford, California 94305, USA}
\affiliation{Kavli Institute for Particle Astrophysics and Cosmology, SLAC National Accelerator Laboratory, 2575 Sand Hill Rd, Menlo Park, California 94025, USA}

\date[Published in PRL 20 January 2016]{}

\begin{abstract}
We present results from an analysis of all data taken by the 
\biceptwo\ \& \keckarray\ CMB polarization experiments
up to and including 
the 2014 observing season.
This includes the first \keckarray\ observations at 95\,GHz.
The maps reach a depth of 50\,nK\,deg in Stokes $Q$ and $U$
in the 150\,GHz band and 127\,nK\,deg in the 95\,GHz band.
We take auto- and cross-spectra between these maps and
publicly available maps from \wmap\ and \planck\ at frequencies
from 23\,GHz to 353\,GHz.
An excess over lensed-\lcdm\ is detected at modest
significance in the 95$\times$150 $BB$ spectrum,
and is consistent with the dust contribution expected
from our previous work.
No significant evidence for synchrotron emission is found
in spectra such as 23$\times$95, or for correlation between the
dust and synchrotron sky patterns in spectra such as 23$\times$353.
We take the likelihood of all the spectra for a
multi-component model including lensed-\lcdm, dust,
synchrotron and a possible contribution from inflationary gravitational waves 
(as parametrized by the tensor-to-scalar ratio $r$),
using priors on the frequency spectral behaviors
of dust and synchrotron emission from previous analyses of
\wmap\ and \planck\ data in other regions of the sky.
This analysis yields an upper limit $r_{0.05}<0.09$ at 95\% confidence,
which is robust to variations explored in analysis and priors.
Combining these \bmode\ results with the (more model-dependent) constraints
from \planck\ analysis of CMB temperature plus BAO and other data, yields a combined
limit $r_{0.05}<0.07$ at 95\% confidence.
These are the strongest constraints to date on inflationary gravitational waves.
\end{abstract}

\doi{xyz}

\maketitle

{\it Introduction.}---Measurements of the cosmic microwave background
(CMB)~\cite{penzias65} are one of the observational pillars
of the standard cosmological model (\lcdm) and constrain
its parameters to high precision
(see most recently Ref.~\cite{planck2015XIII}).
This model extrapolates the Universe back to very high
temperatures ($\gg 10^{12}$\,K) and early times ($\ll 1$\,s).
Observations indicate that conditions at these early times 
are described by an almost uniform
plasma with a nearly scale invariant spectrum of adiabatic
density perturbations.
However, \lcdm\ itself offers no explanation for how these
conditions occurred.
The theory of inflation is an extension to the standard
model, which postulates a phase of exponential expansion at
a still earlier epoch ($\sim10^{-35}$\,s)
that precedes \lcdm\ and produces the required initial conditions
(See Ref.~\cite{kamionkowski2015} for a recent review and 
citations to the original literature.)

There is widespread support for the claim that existing observations
already indicate that some version of inflation probably did occur,
but there are also skeptics~\cite{guth14,ijjas14}.
As well as the specific form of the initial density
perturbations there is an additional relic which inflation predicts,
and which one can attempt to detect. 
Inflation launches tensor mode perturbations 
into the fabric of space-time which will propagate 
unimpeded as inflationary gravitational waves (IGWs) 
to the present day. 
Their amplitude is diminished with the expansion
of the Universe, and detection at the present epoch 
is not feasible with current technology.
The most promising potential method of detection is to
look for their signature written into the pattern
of the CMB at last scattering, 380,000 years after the
Universe entered the realm of fully known physics.
Inflationary theories generically
predict that IGWs exist, but many specific models
have been proposed producing a wide range of amplitudes---with
some being unobservably small~\cite{kamionkowski2015}.
The size of the IGW signal is conventionally expressed
as the initial ratio of the tensor and scalar
perturbation amplitudes $r$.

In the \lcdm\ standard model the CMB is polarized
by Thomson scattering of Doppler induced quadrupoles
in the local radiation field at last scattering.
This naturally produces a polarization pattern
with direction parallel/perpendicular to the gradient
of its intensity---this is curl-free, or $E$-mode
polarization, and was first detected in Ref.~\cite{kovac02}.
Due to small gravitational deflections of the CMB
photons in flight by intervening large scale
structure, the initial purity of the
$E$-mode pattern is disturbed and a small lensing
$B$-mode is produced at sub-degree angular
scales~\cite{polarbear14,keisler15}.

IGWs are
intrinsically quadrupolar distortions of the metric
and produce both $E$ and $B$-mode polarization
depending on their orientation with respect to our
last scattering surface.
However, due to the large \lcdm\ $E$-mode signal,
the most promising place to search for an IGW signal
is in $B$-modes.
Furthermore, since the IGW $B$-modes have a much
redder spectrum than the lensing $B$-modes,
the best place to look is at angular scales larger
than a few degrees (multipoles $\ell < 100$).
Limits on IGW from non-polarized CMB observations
are now fully saturated at cosmic variance
limits~\cite{planck2015XIII} and it is generally agreed
that the best (only) way to make further progress is
through improved measurements of CMB $B$-modes.

The \bicep\ and \keckarray\ telescopes are small
aperture polarimeters specifically designed to
search for an IGW signal at the recombination bump
($\ell \approx 80$).
\bicepone\ operated from 2006 to 2008 and set a limit
$r_{0.05}<0.70$ at 95\% confidence~\citep{barkats14}.
\biceptwo\ operated from 2010 to 2012 at 150\,GHz and
in Ref.~\cite{biceptwoI}
reported a detection of a substantial excess over the
lensed-\lcdm\ expectation in the multipole range $30<\ell<150$.
Additional measurements at 150\,GHz taken by the \keckarray\
during 2012 and 2013 confirmed this excess~\citep{biceptwoV}.
However, new data from the \planck\ space mission provided 
evidence that emission from galactic dust grains could be more
polarized at high galactic latitudes than anticipated~\citep{planckiXIX, planckiXXX},
a possibility emphasized by~\citep{flauger14,mortonson14}.
Analysis of the combined \biceptwo\ and \keckarray\
150\,GHz data in combination with data from \planck\
(principally at 353\,GHz) showed that a substantial
part of the 150\,GHz excess is due to polarized emission from
galactic dust grains, and that once this is accounted
for, the result becomes $r_{0.05}<0.12$ at 95\% confidence~\citep{bkp}.

\biceptwo\ was a simple 26~cm aperture all-cold refractor,
and \keckarray\ is basically five copies of this on a single
telescope mount~\citep{biceptwoII,biceptwoV}.
Both are sited at the South Pole in Antarctica, taking
advantage of the dry atmosphere and stable observing conditions.
In addition to the all-cold optics these telescopes
have two features which aid greatly in the suppression and
characterization of instrumental systematics: i) they are
equipped with co-moving absorptive forebaffles resulting
in extremely low far side-lobe response, and ii) the
entire instrument can be rotated about the line of
sight allowing modulation of polarized signal.

\keckarray\ was designed at the outset to observe
in multiple frequency bands---the 2012 and 2013 
observations were all taken at 150\,GHz
because detectors for other bands were not yet ready.
Before the 2014 season two of the five receivers of \keckarray\
were refitted for operation in a band centered on 95\,GHz
(the other three receivers remaining unchanged at 150\,GHz).
In this paper we fold in this new data and perform
a multi-component, multi-spectral likelihood analysis similar
to our previous analysis~\cite{bkp}.

This paper builds on the initial \biceptwo\ results
paper~\cite[hereafter BK-I]{biceptwoI},
the \keck\ 2012+2013 results paper~\cite[hereafter BK-V]{biceptwoV},
and the \biceptwo/\keck/\planck\ analysis paper~\cite[hereafter BKP]{bkp}.

{\it Instrument and observations.}---The \keckarray\ instrument
is described in Sec.~2 of BK-V.
(See also the \biceptwo\ Instrument Paper~\cite{biceptwoII} for further details.)
Before the 2014 observing season two of the receivers
of \keckarray\ were removed,
the lenses and filters were replaced with versions optimized
for a band centered at 95\,GHz, and the focal planes were replaced
with units loaded with appropriately scaled versions of our
antenna-coupled detectors~\citep{bkdets}.
Because the physical size of these antennas is larger
each of the four tiles contains only a $6\times6$
array (rather than $8\times8$ at 150\,GHz).
With two focal planes at 95\,GHz this gives 288 total
detector pairs (576 total detectors).

During the 2014 austral winter season the array
was operated exactly as for the previous seasons.
A $\sim 1$\% region of sky centered at RA 0h, Dec.\ $-57.5\deg$
was observed from March until November over $\approx 4600$
fifty minute ``scansets''.
Efficiency and yield was similar to previous seasons.
See Sec.~4 of BK-V for further details
of the observing strategy and data selection.

{\it \biceptwo/\keck\ Maps.}---The processing from
time stream to maps is identical
to that described in Sec.~III \&~IV of~BK-I
and summarized in Sec.~5 of~BK-V.
Relative gain calibration is applied between the two
halves of each pair and the difference is taken.
Filtering is then applied to remove residual atmospheric
noise and any ground-fixed (scan-synchronous) pickup.
The data are then binned into simple map pixels and,
with knowledge of the polarization sensitivity directions,
maps of Stokes parameters $Q$ and $U$ are formed.
``Deprojection'' is also performed to remove leakage
of temperature to polarization due to beam systematics and
this results in an additional filtering of signal.

\begin{figure*}
\resizebox{\textwidth}{!}{\includegraphics{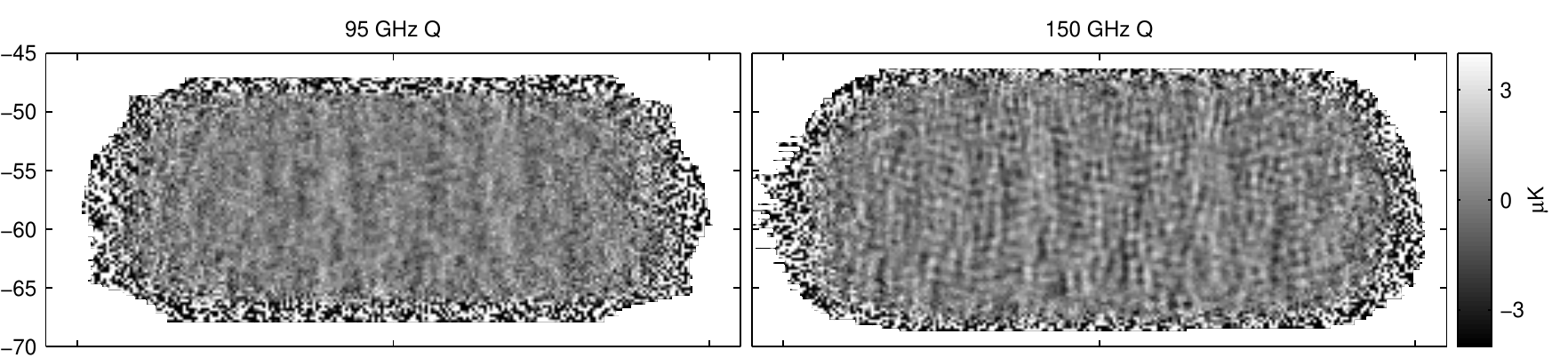}}
\caption{Deep $Q$ maps at 95 \& 150\,GHz
using all \biceptwo/\keck\ data 
through the end of
the 2014 observing season---we refer to these maps as BK14.
Noise levels are 127~nK$\,$deg (left) and 50~nK$\,$deg (right).
Both maps show a high signal-to-noise pattern dominated by
E-mode polarization; the 95\,GHz
maps appear smoother because of the larger beam size.}
\label{fig:qu_maps}
\end{figure*}

Fig.~\ref{fig:qu_maps} shows the 95 \& 150\,GHz
$Q$ maps combining data from \biceptwo\ (2010--2012) and \keckarray\
(2012--2014)---we refer to these as the BK14 maps meaning
that they contain all data up to and including
that taken during the 2014 observing season.
The 150\,GHz maps add 3 more receiver years to the previous 13
in the BK13 based analysis of BKP, and modestly improves the
$Q$/$U$ sensitivity from 57~nK$\,$deg to 50~nK$\,$deg
(3.0~$\mu$K$\,$arcmin) over an effective area of 395 square degrees.
These are the deepest maps of CMB polarization published to date.
The 95\,GHz maps contain only 2 receiver years of data
and the $Q$/$U$ sensitivity is 127~nK$\,$deg
(7.6~$\mu$K$\,$arcmin) over an effective area of 375 square degrees.
(The survey weight is thus 310,000 (47,000)\,$\mu$K$^{-2}$
at 150 (95)\,GHz.)
The 95\,GHz beam is wider (43 arcmin versus 30 arcmin FWHM)
and we see the effect of the additional beam smoothing.
However, the degree scale structure is clearly nearly identical
at both frequencies.
While there is a dust component hidden in the 150\,GHz maps it
is highly subdominant to \lcdm\ \emode\ signal.
See Appendix~A for the full set
of $T$/$Q$/$U$ signal and noise maps.

{\it External Maps.}---
We use the Public Release 2 ``full mission'' maps available from the
\planck\ Legacy Archive~\footnote{See \url{http://www.cosmos.esa.int/web/planck/pla}
}\citep{planck2015I},
noting that these are nearly identical to those used in BKP.
For this analysis we also add the \wmap9 23\,GHz (K-band) and 33\,GHz (Ka-band)
maps~\footnote{See \url{http://lambda.gsfc.nasa.gov/product/map/dr5/m_products.cfm}}\citep{bennett13}.

For each of these external maps we deconvolve the native instrument
beam, reconvolve the Keck 150\,GHz beam, and then process the result
through an ``observing'' matrix to produce a map with the same
filtering of spatial modes as the 150\,GHz map.
See Sec.~II.A of BKP for further details of this process.
For \planck\ we use the FFP8 simulations~\citep{planck2015XII}
and for \wmap\ we use simple inhomogeneous white noise simulations
derived from the provided variance maps.

{\it Power Spectra.}---We convert the maps to power spectra
using the methods described
in Sec.~VI of BK-I including the matrix based purification
operation to prevent $E$ to $B$ mixing.
We generate separate purification matrices to match
the filtering of the 95 \& 150\,GHz maps.

We first subject the new 95\,GHz data to our usual suite of ``jackknife''
internal consistency checks.
The results are given in Appendix~B and show
empirically that the data are free of systematic contamination
at a level greater than the noise.
In addition, in Appendix~C we investigate
the stability of the previous 150\,GHz spectrum when
adding the new 2014 data---there is no indication of problems.

\begin{figure*}
\resizebox{\textwidth}{!}{\includegraphics{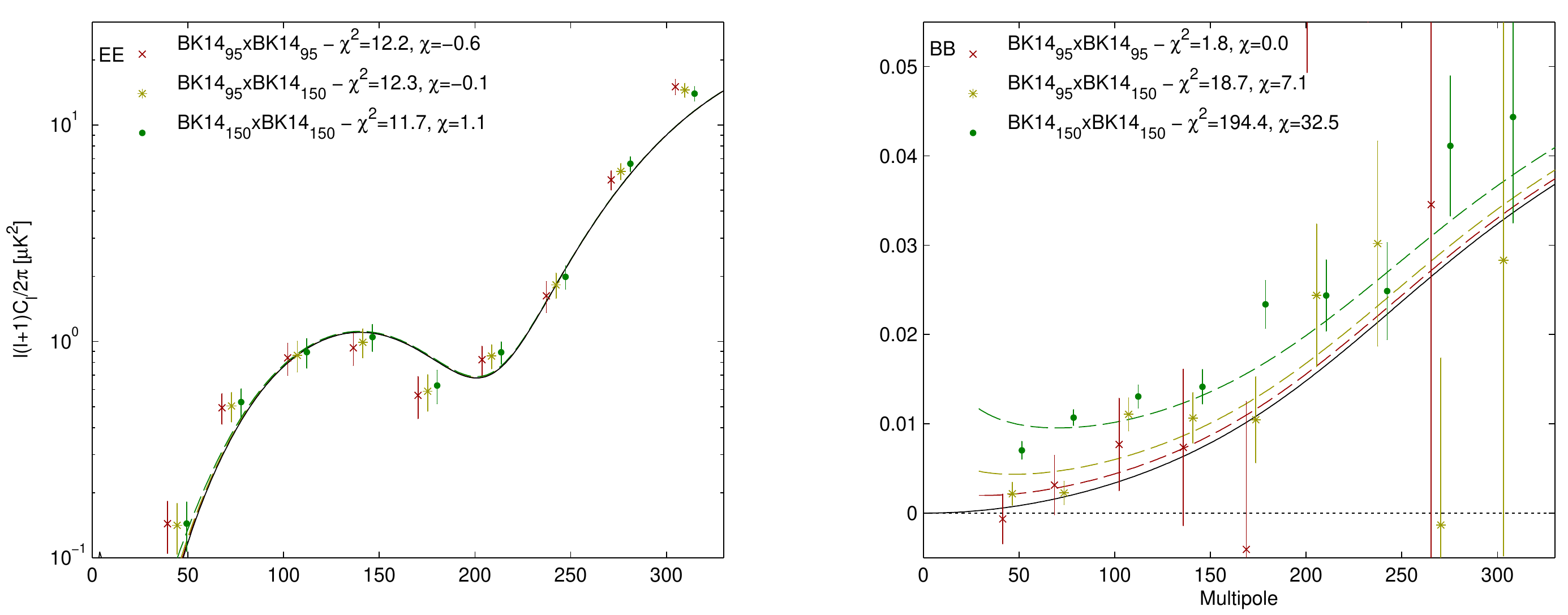}}
\caption{$EE$ and $BB$ auto- and cross-spectra between 95 \& 150\,GHz
using all \biceptwo/\keck\ data up to and including that taken
during the 2014 observing season---we refer to these spectra as BK14. 
(For clarity the sets of points are offset horizontally.)
The solid black curves show the lensed-\lcdm\ theory spectra.
The error bars are the standard deviations of the
lensed-\lcdm+noise simulations and hence contain no sample variance
on any additional signal component.
The $\chi^2$ and $\chi$ (sum of deviations) against
lensed-\lcdm\ for the lowest five bandpowers are given,
and can be compared to their expectation value/standard-deviation
of 5/3.1 and 0/2.2 respectively
The dashed lines show a
lensed-\lcdm+dust model derived from our previous BKP analysis.}
\label{fig:powspecres_95x150}
\end{figure*}

We now proceed to comparing the spectra and cross spectra
of our 95 and 150\,GHz maps---Fig.~\ref{fig:powspecres_95x150}
shows the results.
We use a common apodization mask as the geometric mean of
the two (smoothed) inverse variance maps.
The $EE$ spectra agree to within much better than the nominal
error bar size because the uncertainty is dominated by
sample variance and we are observing the same piece of
sky.
To make a rough estimate of the significance of deviation
from lensed-\lcdm, we calculate $\chi^2$ and $\chi$
(sum of normalized deviations) as shown on the plot.
We see strong evidence for excess $BB$ power in BK14$_{150}\times$BK14$_{150}$
and moderate evidence in BK14$_{95}\times$BK14$_{150}$.
Dashed lines for the lensed-\lcdm+dust model derived in BKP
are over-plotted and appear to be consistent with the new data.

\begin{figure}
\begin{center}
\resizebox{\columnwidth}{!}{\includegraphics{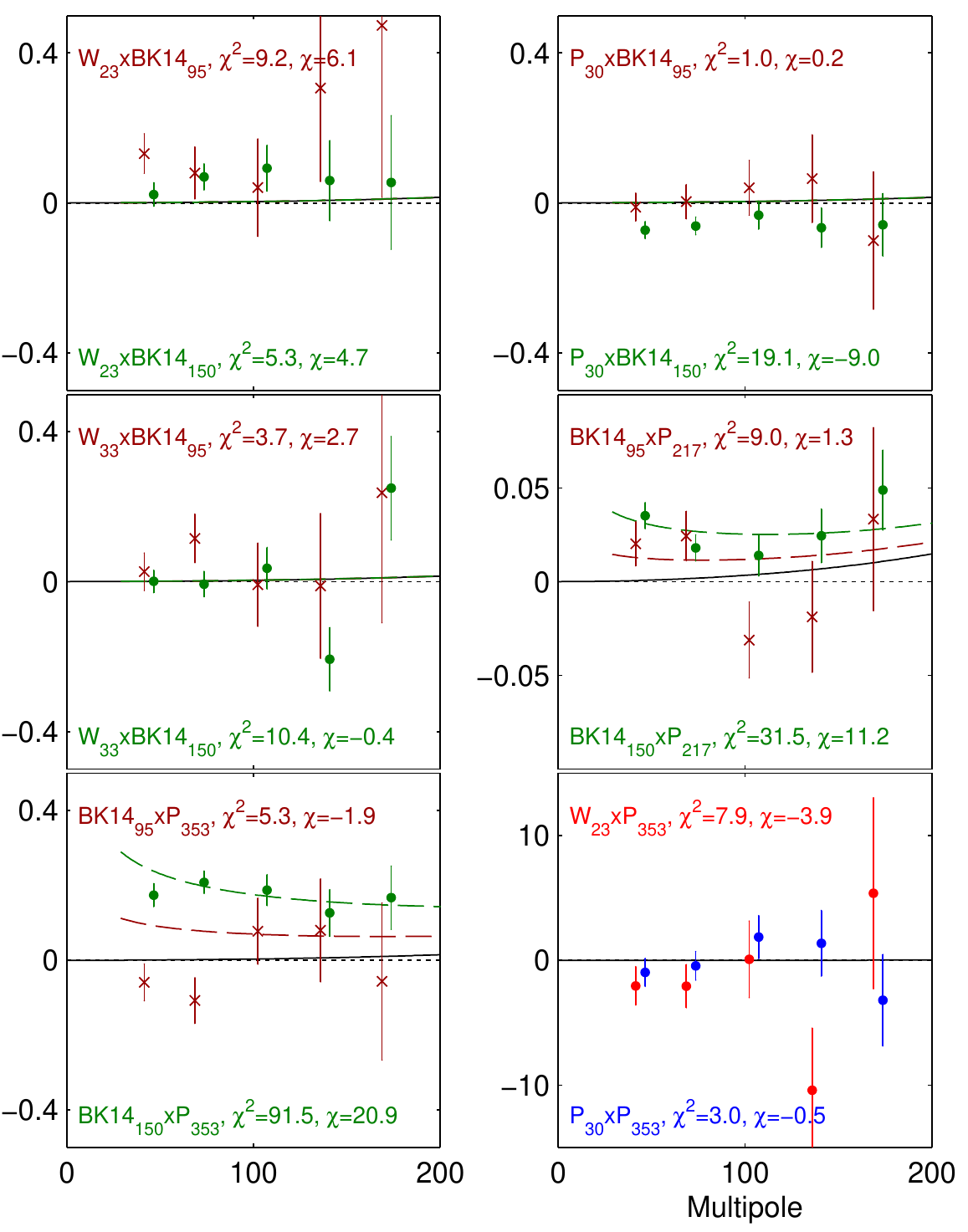}}
\end{center}
\caption{Selected $BB$ cross-spectra between the BK14 maps
at 95 (red) \& 150\,GHz (green) and the external maps of \wmap\ and \planck. 
The quantity plotted is $\clstar$ ($\mu$K$^2$),
and the error bars are the standard deviations of the
lensed-\lcdm+noise simulations.
The solid black curves show the lensed-\lcdm\ theory spectrum
and the $\chi^2$ and $\chi$ versus this model are shown.
W$_{23}\times$BK14$_{95}$ and W$_{23}\times$BK14$_{150}$ are both
mildly elevated showing weak evidence for synchrotron but
P$_{30}\times$BK14$_{150}$ has stronger nominal anticorrelation.
We see modest evidence for detection of dust emission in
BK14$_{150}\times$P$_{217}$ and strong evidence in BK14$_{150}\times$P$_{353}$.
The dashed lines show a
lensed-\lcdm+dust model derived from our previous BKP analysis.}
\label{fig:powspec_BKxExt}
\end{figure}

Fig.~\ref{fig:powspec_BKxExt} shows selected $BB$ cross
spectra between the BK14 95 \& 150\,GHz maps and the
\planck\ (P) and \wmap\ (W) bands.
There is no strong evidence for detection of synchrotron
emission---W$_{23}\times$BK14$_{95}$ and W$_{23}\times$BK14$_{150}$ are both
mildly elevated but P$_{30}\times$BK14$_{150}$ has stronger nominal
anticorrelation (as noted in the BKP paper).
W$_{33}\times$BK14$_{95}$ and W$_{33}\times$BK14$_{150}$ are both consistent with null.
The only strong detections of excess signal are
in BK14$_{150}\times$P$_{353}$ and, at lower significance BK14$_{150}\times$P$_{217}$.
See Appendix~D for the full set of auto- and cross-spectra.

{\it Likelihood Analysis.}---We next proceed to a multicomponent,
multi-spectral likelihood
analysis which is an expanded version of that described
in Sec.~III of the BKP paper.
We compute the likelihood of the data for any given
proposed model using an extended version of the
HL approximation~\citep{hamimeche08} and
the full covariance matrix of the
auto- and cross-spectral bandpowers as derived from simulations
(setting to zero terms whose expectation value is zero).

In this analysis we primarily use a
lensed-\lcdm+dust+synchrotron+$r$ model and 
explore the parameter space using \texttt{COSMOMC}~\citep{cosmomc}.
The \texttt{COSMOMC} module containing the data and model
is available for download at \url{http://bicepkeck.org}.
In this paper the ``baseline'' analysis is defined to:

\begin{itemize}
\item Use the BK14 maps as shown in Fig.~\ref{fig:qu_maps}
(all \biceptwo/\keck\ data up to and including that taken
during the 2014 observing season).
\item Use all the polarized bands of \planck\ (30--353\,GHz) plus
the 23 \& 33\,GHz bands of \wmap.
\item Use all possible $BB$ auto- and cross-spectra between these maps.
This includes all the spectra shown in Figures~\ref{fig:powspecres_95x150}
\&~\ref{fig:powspec_BKxExt}---the complete set are shown in Appendix D.
Spectra with no detection can, of course, still have constraining
power---for instance non-detection
in P$_{30}\times$P$_{353}$ disfavors sync/dust correlation.
\item Use nine bandpowers spanning the range $20<\ell<330$.
\item Include dust with amplitude $\Adf$ evaluated
at 353\,GHz and $\ell=80$.
As in the BKP analysis the frequency spectral behavior is taken
as a simple modified black body spectrum with $T_\mathrm{d}=19.6$\,K
and $\Bd=1.59 \pm 0.11$, using a Gaussian prior
with the given $1\sigma$ width.
Analyzing polarized emission at intermediate galactic latitudes
Fig.~11 of Ref.~\cite{planckiXXII} shows that this model is
accurate in the mean to within a few percent over the frequency
range 100--353\,GHz, while the patch-to-patch fluctuation
is noise dominated.
The spatial power spectrum is taken as a simple power law
$\mathcal{D}_\ell \propto \ell^{\ad}$ marginalizing
over the range $-1<\ad<0$,
where $\mathcal{D}_\ell \equiv \clstar$.
\item Include synchrotron with amplitude $\Asf$
evaluated at 23\,GHz (the lowest \wmap\ band) and $\ell=80$,
assuming a simple power law for the frequency spectral behavior
$\As \propto \nu^{\Bs}$ with a Gaussian prior
$\Bs=-3.1\pm0.3$~\citep{fuskeland14}.
The spatial power spectrum is taken as a simple power law
$\mathcal{D}_\ell \propto \ell^{\as}$ marginalizing
over the range $-1<\as<0$.
\item Allow sync/dust correlation and marginalize over the correlation
parameter $0<\epsilon<1$.
\item Quote the tensor/scalar power ratio $r$ at a pivot
scale of 0.05~Mpc$^{-1}$ and fix the tensor spectral index $n_t=0$.
\end{itemize}
See Appendix~E1 for a more detailed
explanation of these choices.

\begin{figure*}
\begin{center}
\resizebox{0.7\textwidth}{!}{\includegraphics{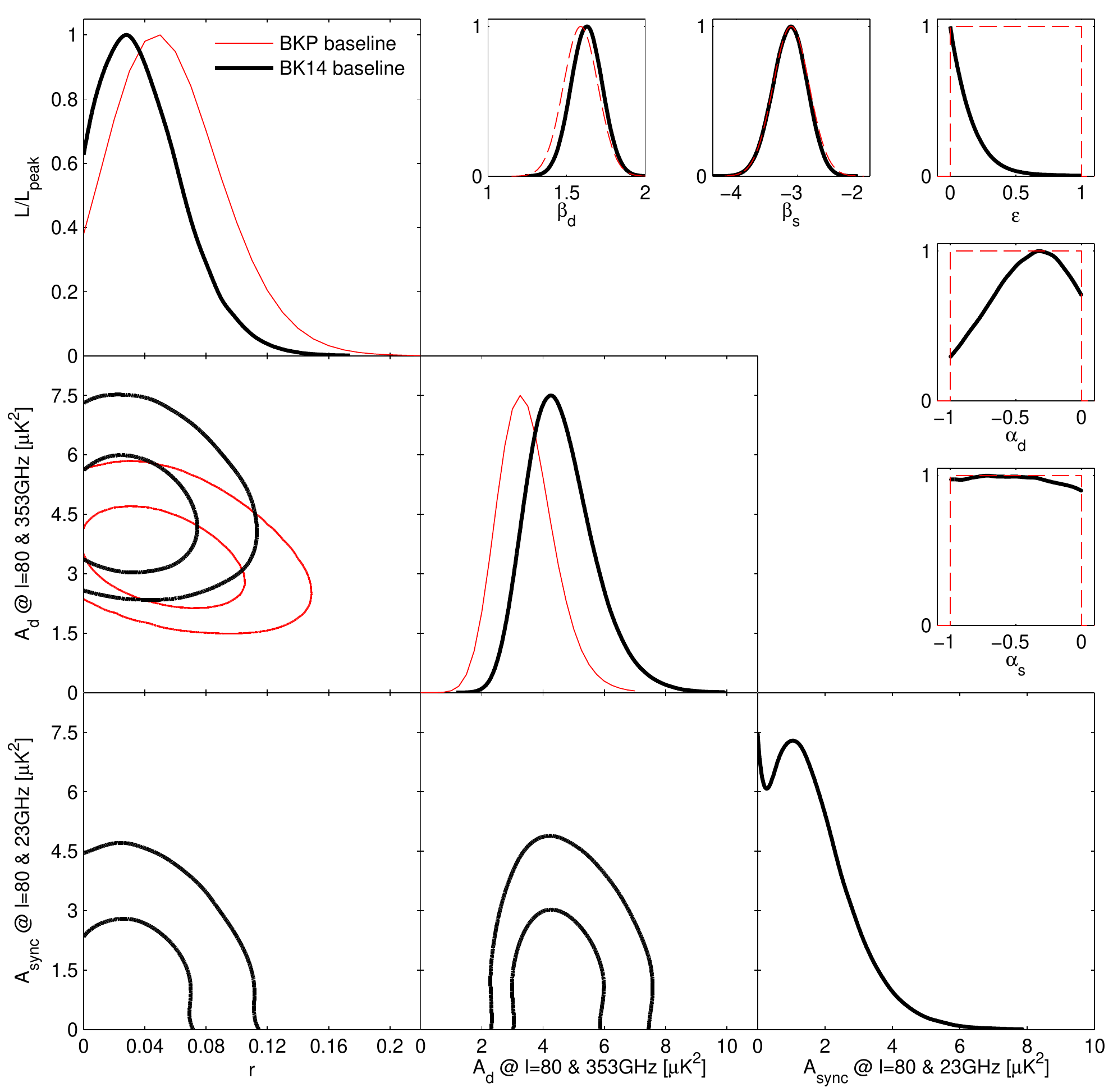}}
\end{center}
\caption{Results of a multicomponent multi-spectral likelihood
analysis of \biceptwo/\keck+external data.
The red faint curves are the primary result from the previous
BKP paper (the black curves from Fig.~6 of that paper).
The bold black curves are the new baseline BK14 results.
Differences between these analyses include adding
synchrotron to the model, including additional external
frequency bands from \wmap\ \& \planck,
and adding \keckarray\ data taken
during the 2014 observing season at 95 \& 150\,GHz.
We see that the peak position of the tensor/scalar ratio
curve $r$ shifts down slightly and the upper limit tightens
to $r_{0.05}<0.09$ at 95\protect\% confidence.
The parameters $\Ad$ and $\As$ are the
amplitudes of the dust and synchrotron $B$-mode power spectra, where
$\beta$ and $\alpha$ are the respective frequency and spatial
spectral indices.
The correlation coefficient between the dust and synchrotron patterns is
$\epsilon$.
In the $\beta$, $\alpha$ and $\epsilon$ panels the dashed red lines
show the priors placed on these parameters (either Gaussian or uniform).}
\label{fig:likebase}
\end{figure*}

Results of this baseline analysis are shown in
Fig.~\ref{fig:likebase} and yield the following
statistics:
$r_{0.05}=0.028^{+0.026}_{-0.025}$, $r_{0.05}<0.090$ at 95\% confidence,
$\Adf=4.3^{+1.2}_{-1.0}$\,$\mu$K$^2$, and
$\Asf<3.8$\,$\mu$K$^2$ at 95\% confidence.
For $r$ the zero-to-peak likelihood ratio is 0.63.
Taking
$\frac{1}{2} \left( 1-f \left( -2\log{L_0/L_{\rm peak}} \right) \right)$,
where $f$ is the $\chi^2$ CDF (for one degree of freedom),
we estimate that the probability to get a likelihood ratio smaller than this is
18\% if, in fact, $r=0$.
Running the analysis on the lensed-\lcdm+dust+noise simulations
produces a similar number.
The zero-to-peak likelihood ratio for $\Ad$ indicates that the
detection of dust is now $>8\sigma$.

Results for the additional parameters are shown in the
upper right part of Fig.~\ref{fig:likebase}.
The dust frequency spectral parameter $\Bd$
pulls weakly against the prior to higher values.
The synchrotron frequency spectral parameter $\Bs$
just reflects the prior (as expected since synchrotron
is not strongly detected).
The data have a mild preference for
values of $\ad$ close to the $-0.42$ found
in Ref.~\cite{planckiXXX}, while $\as$ is
unconstrained.
The data disfavor strong sync/dust
correlation (due to the non detection of signal in
spectra like $W_{23}\times$P$_{353}$---see
Fig~\ref{fig:powspec_BKxExt}).
As $\As$ approaches zero $\epsilon$ becomes
unconstrained leading to an increase in the available
parameter volume, and the ``flare'' in the $\As$ constraints.

The maximum likelihood model (including priors) has parameters
$r_{0.05}=0.026$, $\Adf=4.1$\,$\mu$K$^2$,
$\Asf=1.4$\,$\mu$K$^2$, 
$\Bd=1.6$, $\Bs=-3.1$,
$\ad=-0.19$, $\as=-0.56$,
and $\epsilon=0.00$.
This model appears to be an acceptable fit to the data---see
Appendix~D for further details.

\begin{figure}
\begin{center}
\resizebox{0.7\columnwidth}{!}{\includegraphics{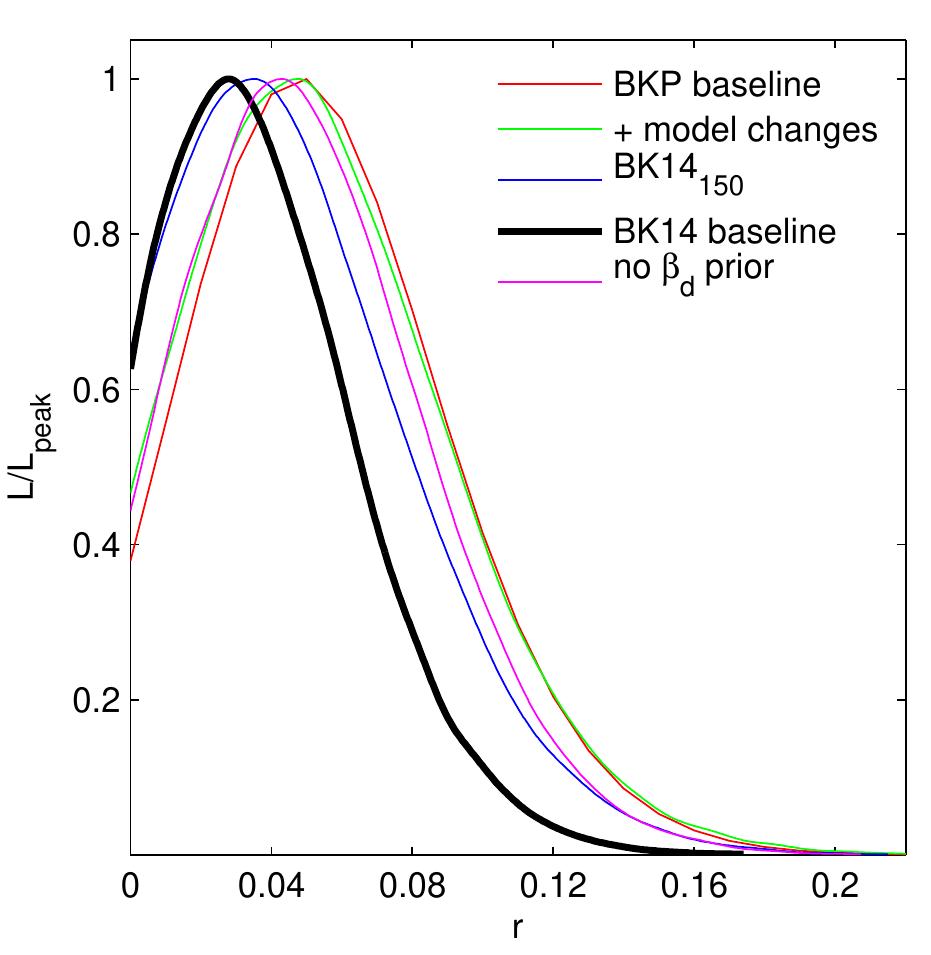}}
\end{center}
\caption{Likelihood results on $r$ for several intermediate
steps between the BKP (previous) and BK14 (current) analyses.
See text for details.}
\label{fig:evolmain}
\end{figure}

In Fig.~\ref{fig:likebase} we see that as compared to the
primary BKP analysis the peak position of the likelihood curve
for $r$ has shifted down slightly.
In Fig.~\ref{fig:evolmain} we investigate why.
Although we have made extensive changes to the
model, these make only a small difference.
(See Appendix~E1 for details
of these changes.)
The change from the BK13$_{150}$ to the
BK14$_{150}$ maps causes some of the downward shift
in the peak position.
This may seem surprising given that only a relatively
small amount of additional data has been added ($\sim20$\%).
However Appendix~C shows that the shifts in
the bandpower values are not unlikely and
we should therefore accept the shift in the $r$ constraint
as simply due to noise fluctuation.
Adding in the BK14$_{95}$ data produces an additional
downward shift in the peak position, and
also significantly narrows the likelihood curve.

Fig.~\ref{fig:evolmain} shows one additional variation.
It turns out that the tight prior on $\Bd$
from \planck\ analysis of other regions of sky
is becoming unnecessary.
Removing the prior the peak position of the
likelihood on $r$ shifts
up slightly and broadens so that $r_{0.05}=0.043^{+0.033}_{-0.031}$
\& $r_{0.05}<0.11$ (95\%), while the likelihood curve
for $\Bd$ is close to Gaussian in shape with mean/$\sigma$ of 1.82/0.26.
In Appendix~E2 we investigate a variety
of other variations from the baseline analysis and in
Appendix~E3 we perform some validation
tests of the likelihood using simulations.

\begin{figure}
\begin{center}
\resizebox{\columnwidth}{!}{\includegraphics{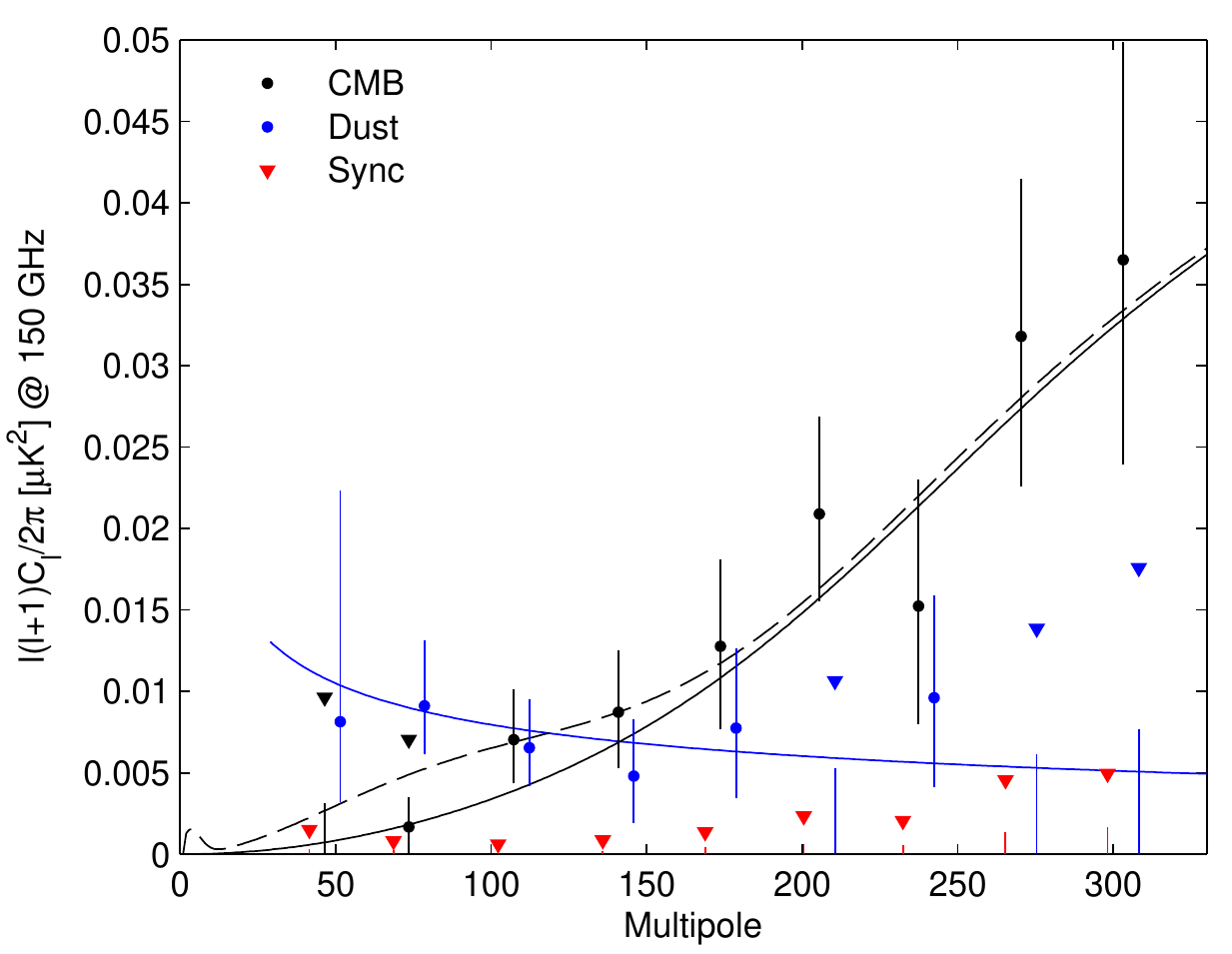}}
\end{center}
\caption{Spectral decomposition of the $BB$ data into synchrotron
(red), CMB (black) and dust (blue) components.
The decomposition is calculated independently in each bandpower,
marginalizing over $\Bd$, $\Bs$, and $\epsilon$ with
the same priors as the baseline analysis.
Error bars denote 68\protect\% credible intervals, with the point marking the most
probable value.
If the 68\protect\% interval includes zero, we also indicate the 95\protect\% upper
limit with a downward triangle.
(For clarity the sets of points are offset horizontally.)
The solid black line shows lensed-\lcdm\ with the dashed
line adding on top an $r_{0.05}=0.05$ tensor contribution.
The blue curve shows a dust model consistent with the
baseline analysis ($\Adf=4.3$\,$\mu$K$^2$,
$\Bd=1.6$, $\ad=-0.4$).}
\label{fig:specdecomp}
\end{figure}

For the purposes of presentation
we also run a likelihood analysis
to find the CMB and foreground contributions on
a bandpower-by-bandpower basis.
The baseline analysis is a single fit to all 9 bandpowers
across 66 spectra with 8 parameters.
Instead we now perform 9 separate
fits---one for each bandpower---across the 66 spectra,
with 6 parameters in each fit.
These 6 parameters are the amplitudes of CMB,
dust and synchrotron plus $\Bd$, $\Bs$, and $\epsilon$ with
identical priors to the baseline analysis. 
The results are shown in Fig.~\ref{fig:specdecomp}---the
resulting CMB bandpowers are consistent with lensed-\lcdm\
while the dust bandpowers are consistent with the level
of dust found in the baseline analysis.
Synchrotron is tightly limited in all the bandpowers.

\begin{figure}
\begin{center}
\resizebox{\columnwidth}{!}{\includegraphics{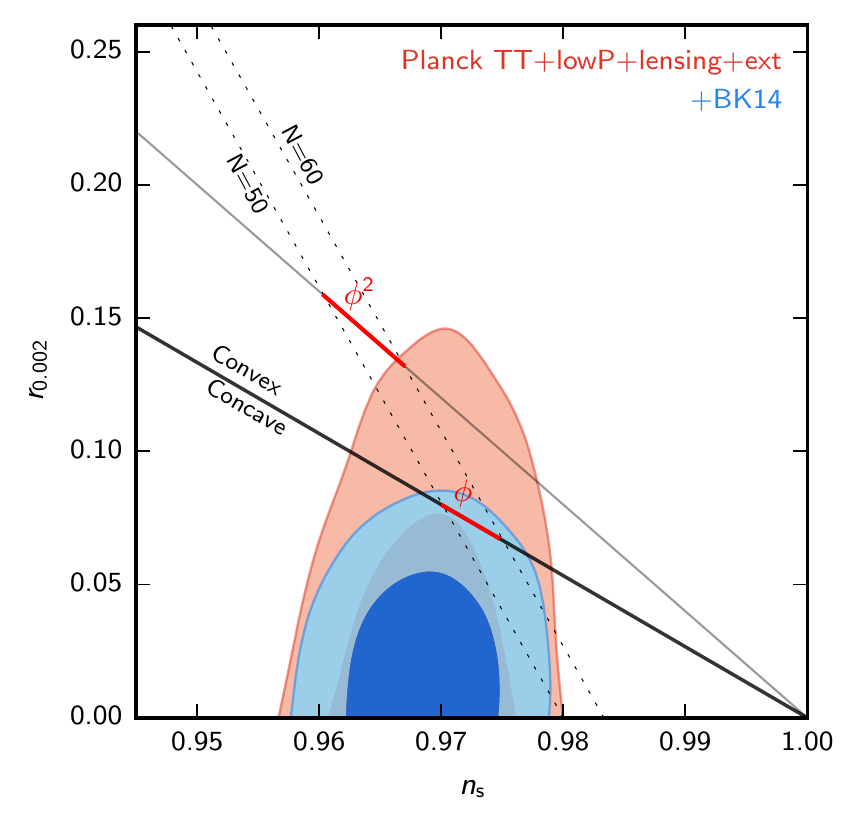}}
\end{center}
\caption{Constraints in the $r$ vs.\ $n_s$ plane when using \planck\
plus additional data,
and when also adding \biceptwo/\keck\ data through the end of the 2014
season including new 95~GHz maps---the constraint on $r$ tightens
from $r_{0.05}<0.12$ to $r_{0.05}<0.07$.
This figure is adapted from Fig.~21 of Ref.~\cite{planck2015XIII}---see
there for further details.}
\label{fig:rns}
\end{figure}

{\it Conclusions.}---As shown above,
the BK14 data in combination with external maps produce
\bmode\ based constraints on the tensor-to-scalar ratio $r$
which place an upper limit $r_{0.05}<0.09$ at 95\% confidence.
The analysis of \planck\ full mission $TT$ data
in conjunction with external data
produces the constraint $r_{0.002}<0.11$ ($r_{0.05}<0.12$) 
at 95\% confidence
(``\planck\ $TT$+lowP+lensing+ext'' in Equation 39b of
Ref.~\cite{planck2015XIII}), and are saturated at cosmic
variance limits.  The BK14 result constitutes the 
first \bmode\ constraints that clearly surpass those from
temperature anisotropies.
In Fig.~\ref{fig:rns} we reproduce Ref.~\cite{planck2015XIII}'s result
in the $r$ vs.\ $n_s$ plane,
and show the effect of adding in our BK14 \bmode\ data.
The allowed region tightens and the joint
result is $r_{0.05}<0.07$ (95\%), 
although as emphasized in Ref.~\cite{planck2015XIII} the
$TT$ derived constraints on $r$ are more model dependent than
$BB$ ones.

\begin{figure}
\resizebox{\columnwidth}{!}{\includegraphics{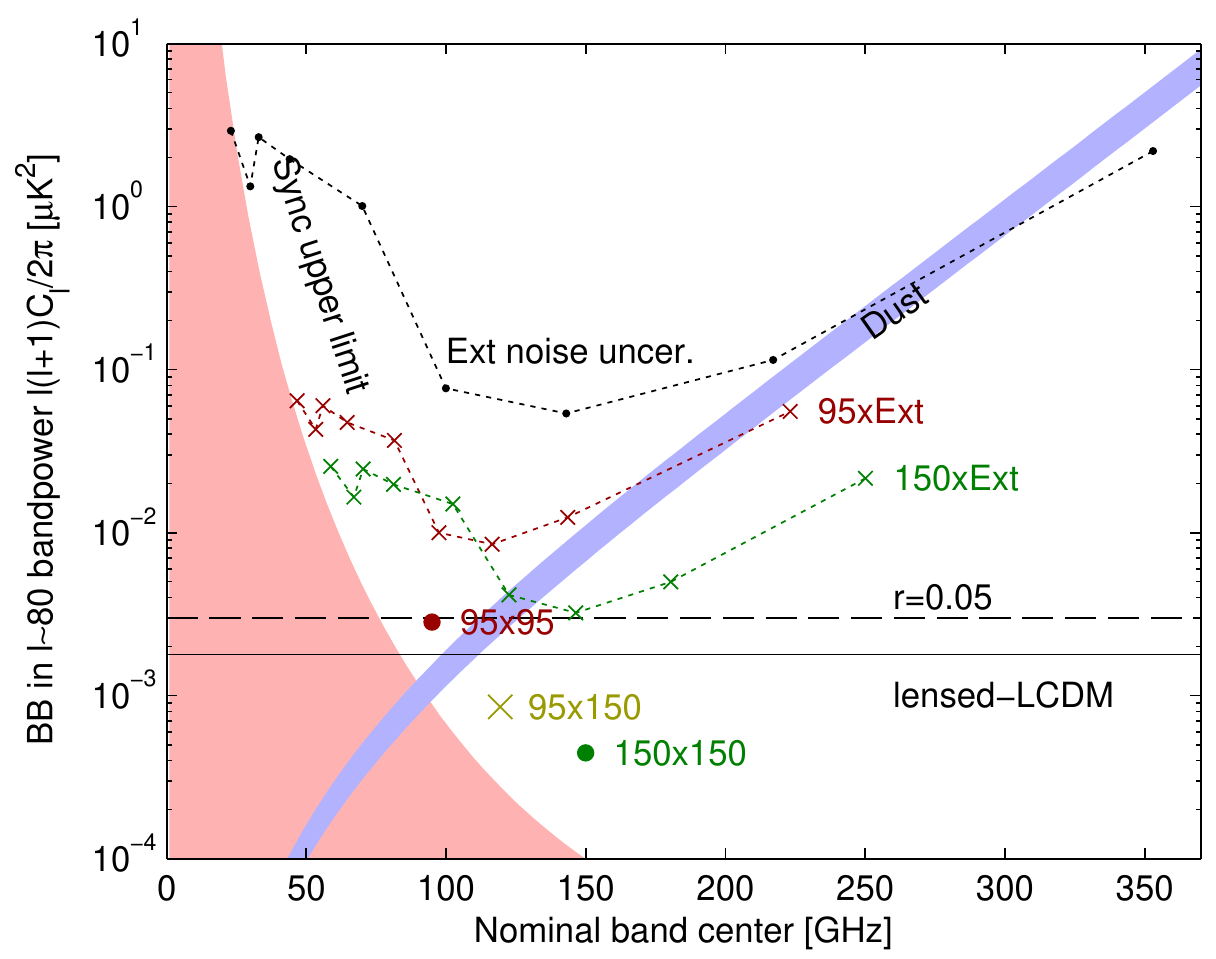}}
\caption{
Expectation values and noise uncertainties
for the $\ell\sim80$
$BB$ bandpower in the \biceptwo/\keck\ field.
The solid and dashed black lines show the expected signal power of
lensed-\lcdm\ and $r_{0.05}=0.05$. 
Since CMB units are used, the levels corresponding
to these are flat with frequency.
The blue band shows a dust model consistent with the baseline analysis
($\Adf=4.3^{+1.2}_{-1.0}$\,$\mu$K$^2$, $\beta_d=1.6$) while
the pink shaded region shows the allowed region for synchrotron
($\Asf<3.8$\,$\mu$K$^2$, $\Bs=-3.1$).
The \biceptwo/\keck\ noise uncertainties are shown as large colored
symbols, and the noise uncertainties of the \wmap/\planck\ single-frequency
spectra evaluated in the \biceptwo/\keck\ field are shown in black.
The red (green) crosses show the noise uncertainty of the
cross-spectra taken between 95 (150)\,GHz and,
from left to right, 23, 30, 33, 44, 70, 100, 143, 217 \& 353\,GHz,
and are plotted at horizontal positions such that they
can be compared vertically with the dust and sync curves.}
\label{fig:noilev}
\end{figure}

Fig.~\ref{fig:noilev} compares signal levels
and current noise uncertainties in the critical $\ell\sim80$ bandpower
(updated from Fig.~13 of BKP).
A second season of 95\,GHz \keckarray\ data has already been recorded
(in 2015) and will push the $95\times95$
point down by a factor of 2.
During 2015 two receivers were also operated
in a third band centered on 220\,GHz,
producing deep maps which will improve dust separation.
This 2015 data is under analysis and will be reported
on in a future paper.
In addition, \bicepthree\ began operations in 2015
in the 95\,GHz band.

In this paper, we have presented an analysis of
all \biceptwo/\keck\ data up through the 2014 season, 
adding, for the first time, 95\,GHz data
from the \keckarray.
We have updated our multi-frequency likelihood analysis with
a more extensive foreground parameterization and the inclusion of
external data from the 23 \& 33~GHz bands of \wmap, 
in addition to all seven polarized bands of \planck.
The baseline analysis yields
$r_{0.05}=0.028^{+0.026}_{-0.025}$ and $r_{0.05}<0.09$ at 95\% confidence,
constraints that are robust to the variations explored in analysis and priors.
With this result, $B$-modes now offer the most powerful limits
on inflationary gravitational waves, surpassing constraints 
from temperature anisotropies and other evidence for the first time.  
With upcoming multifrequency data the \bmode\ constraints 
can be expected to steadily improve.

\acknowledgments

The \keckarray\ project has been made possible through
support from the National Science Foundation under Grants
ANT-1145172 (Harvard), ANT-1145143 (Minnesota)
\& ANT-1145248 (Stanford), and from the Keck Foundation
(Caltech).
The development of antenna-coupled detector technology was supported
by the JPL Research and Technology Development Fund and Grants No.\
06-ARPA206-0040 and 10-SAT10-0017 from the NASA APRA and SAT programs.
The development and testing of focal planes were supported
by the Gordon and Betty Moore Foundation at Caltech.
Readout electronics were supported by a Canada Foundation
for Innovation grant to UBC.
The computations in this paper were run on the Odyssey cluster
supported by the FAS Science Division Research Computing Group at
Harvard University.
The analysis effort at Stanford and SLAC is partially supported by
the U.S. DoE Office of Science.
We thank the staff of the U.S. Antarctic Program and in particular
the South Pole Station without whose help this research would not
have been possible.
Most special thanks go to our heroic winter-overs Robert Schwarz
and Steffen Richter.
We thank all those who have contributed past efforts to the \bicep--\keckarray\
series of experiments, including the \bicepone\ team.
We also thank the \planck\ and \wmap\ teams for the use of their
data.

\bibliography{ms}

\clearpage

\begin{appendix}

\section{Maps}
\label{app:maps}

\begin{figure*}
\resizebox{\textwidth}{!}{\includegraphics{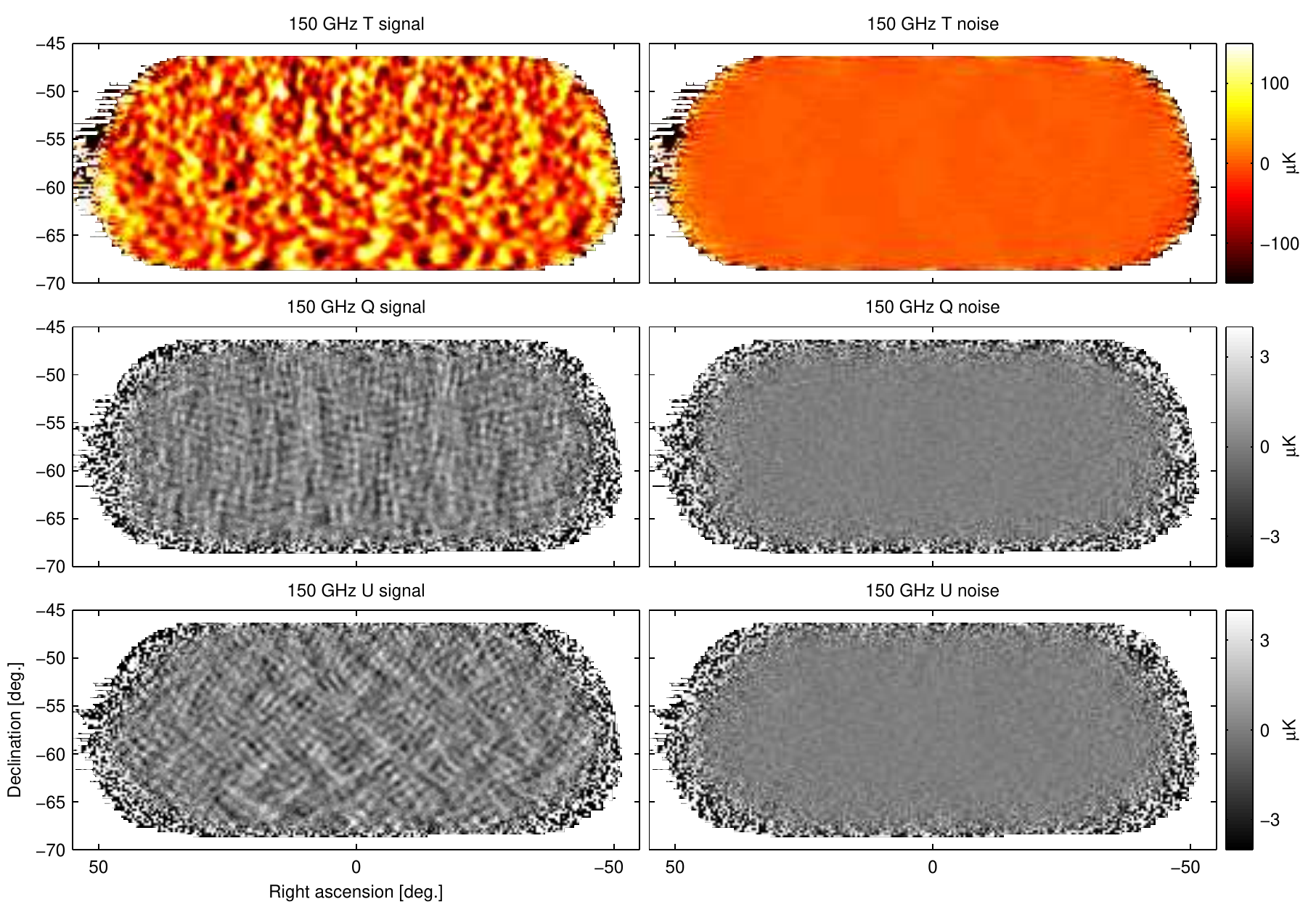}}
\caption{$T$, $Q$, $U$ maps at 150\,GHz
using all \biceptwo/\keck\ data up to and including that taken
during the 2014 observing season---we refer to these maps as BK14$_{150}$.
The left column shows the basic signal maps with $0.25\deg$ pixelization
as output by the reduction pipeline.
The right column shows a noise realization made by randomly assigning
positive and negative signs while coadding the data.
These maps are filtered by the instrument beam (FWHM 30~arcmin),
timestream processing, and (for $Q$ \& $U$) deprojection of
beam systematics.
Note that the horizontal/vertical and $45\deg$ structures seen in the
$Q$ and $U$ signal maps are expected for an \emode\ dominated sky.}
\label{fig:tqu_maps_150}
\end{figure*}

\begin{figure*}
\resizebox{\textwidth}{!}{\includegraphics{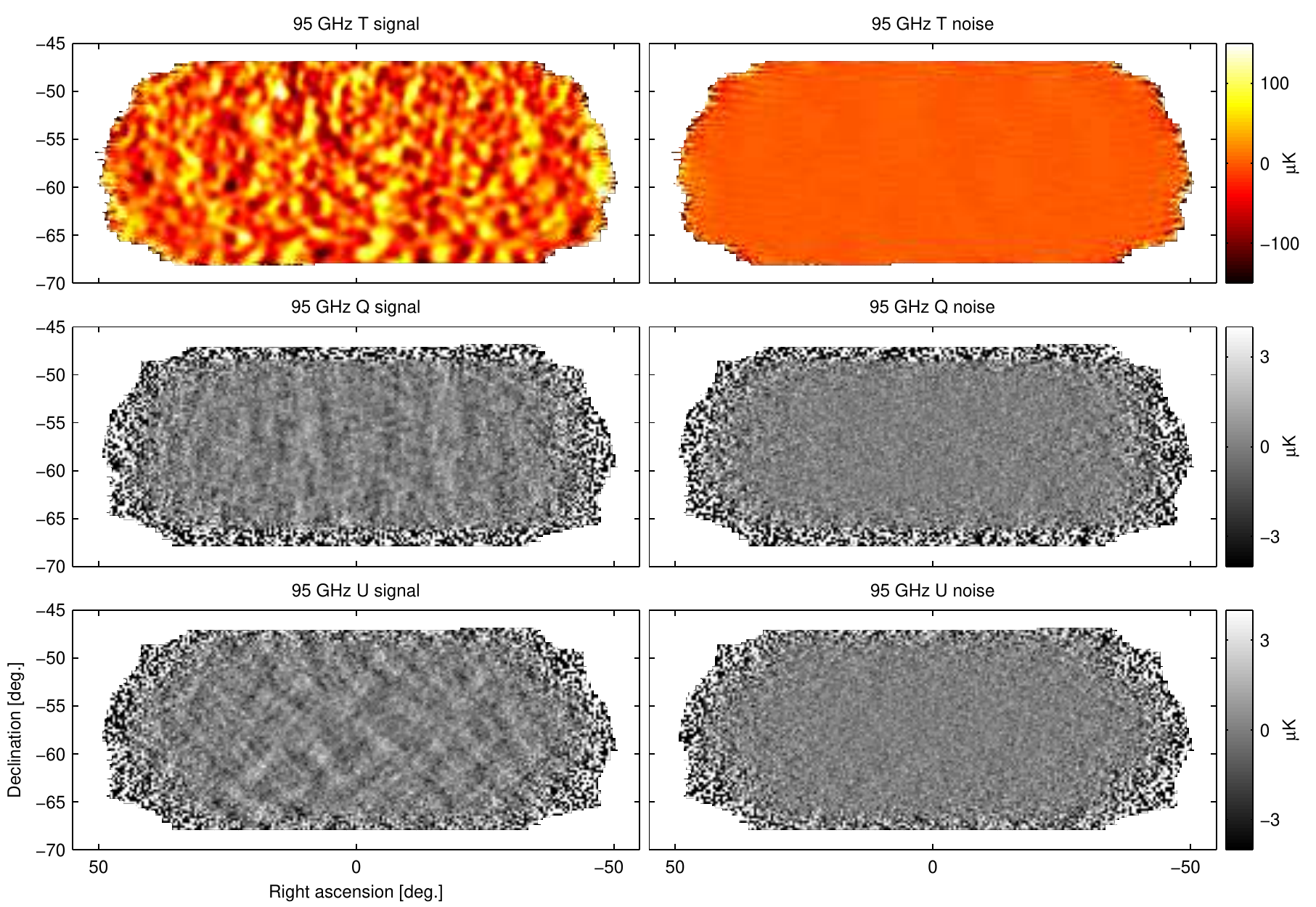}}
\caption{$T$, $Q$, $U$ maps at 95\,GHz
using data taken by two receivers of \keckarray\ during the 2014
season---we refer to these maps as $BK14_{95}$.
These maps are directly analogous to the 150\,GHz maps shown in
Fig.~\ref{fig:tqu_maps_150} except that the
instrument beam filtering is in this case 43~arcmin FWHM.}
\label{fig:tqu_maps_95}
\end{figure*}

Figures~\ref{fig:tqu_maps_150}~\&~\ref{fig:tqu_maps_95}
show the full sets of $T$/$Q$/$U$ maps at 150 \& 95~GHz.
The right side of each figure shows realizations of noise
created by randomly flipping the sign of data subsets while
coadding the map---see Sec.~V.B of~BK-I for
further details.

\section{\keckarray\ 95\,GHz Power Spectra and Internal Consistency Tests}
\label{app:95}

\begin{figure*}[htb]
\resizebox{\textwidth}{!}{\includegraphics{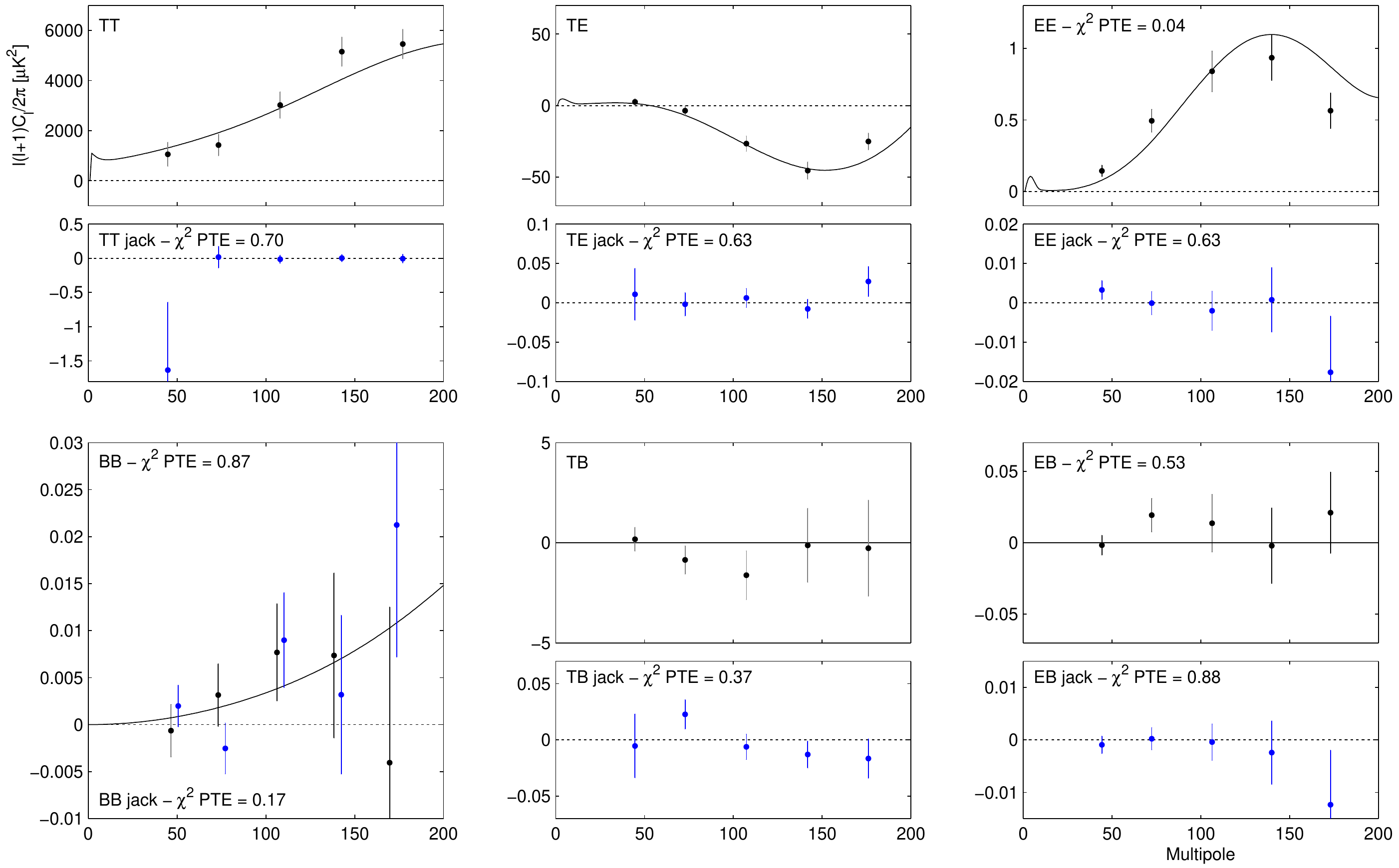}}
\caption{\keckarray\ power spectrum at 95\,GHz for signal (black points)
and deck rotation jackknife (blue points).
The solid black curves show the lensed-\lcdm\ theory spectra.
The error bars are the standard deviations of the
lensed-\lcdm+noise simulations and hence contain no sample variance
on any additional signal component.
The probability to exceed (PTE) the observed value of a simple
$\chi^2$ statistic is given (as evaluated against the simulations).
Note the very different $y$-axis scales for the jackknife spectra
(other than $BB$).
(Also note that the calibration procedure uses $EB$ to set the overall polarization angle
so $TB$ and $EB$ as plotted above cannot be used to measure astrophysical
polarization rotation.)
This figure is analogous to Fig.~2 of BK-I and Fig.~4 of BK-V.}
\label{fig:powspecres_95}
\end{figure*}

A powerful internal consistency test are data split difference tests
which we refer to as ``jackknifes''.
As well as the full coadd signal maps we also form many pairs
of split maps where the splits are chosen such that one might
expect different systematic contamination in the two halves
of the split.
The split halves are differenced and the power spectra
taken.
We then take the deviations of these from the mean of
signal+noise simulations and
form $\chi^2$ and $\chi$ (sum of deviations) statistics.
In this section we perform tests of the new 95\,GHz
data set which are exactly analogous to the tests
of the previous 150\,GHz data sets performed
in Sec.~VII.C of~BK-I and Sec.~6.3 of BK-V.
Fig.~\ref{fig:powspecres_95} shows the signal spectra
and a sample set of jackknife spectra. 
All the signal spectra are consistent with lensed-\lcdm\
and the jackknife spectra with null.

\begin{figure}[htb]
\begin{center}
\resizebox{0.7\columnwidth}{!}{\includegraphics{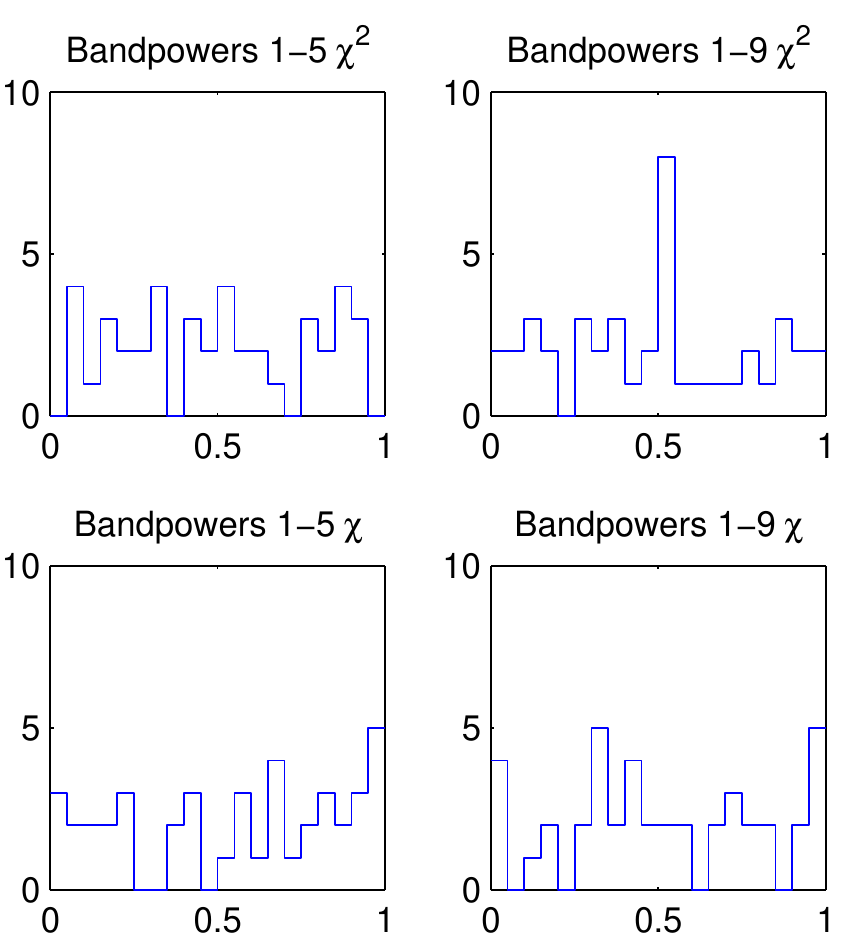}}
\end{center}
\caption{Distributions of the jackknife $\chi^2$ and $\chi$ PTE
values for the \keckarray\ 2014 95\,GHz data over the tests and
spectra given in Table~\ref{tab:ptes}.
This figure is analogous to Fig.~4 of BK-I and Fig.~6 of BK-V.}
\label{fig:ptedist}
\end{figure}

Table~\ref{tab:ptes} shows the $\chi^2$ and $\chi$ statistics
for the full set of 95\,GHz jackknife tests and Fig.~\ref{fig:ptedist}
presents the same results in graphical form.
Note that these values are partially correlated---particularly
the 1--5 and 1--9 versions of each statistic.
We conclude that there is no evidence for corruption of the
data at a level exceeding the noise.

\begingroup
\squeezetable
\begin{table}[pht]
\caption{\label{tab:ptes} Jackknife PTE values from $\chi^2$ and $\chi$ (sum of deviations) tests 
for \keckarray\ 95\,GHz data taken in 2014.
This table is analogous to Table~I of BK-I and Table~4 of BK-V.}
\begin{ruledtabular}
\begin{tabular}{l c c c c }
Jackknife & Band powers & Band powers & Band powers & Band powers\\
& 1--5 $\chi^2$ & 1--9 $\chi^2$ & 1--5 $\chi$ & 1--9 $\chi$ \\
\hline
\\ 
\multicolumn{5}{l}{Deck jackknife} \\ 
EE & 0.625 & 0.591 & 0.523 & 0.569 \\ 
BB & 0.166 & 0.192 & 0.076 & 0.020 \\ 
EB & 0.876 & 0.539 & 0.814 & 0.445 \\ 
\multicolumn{5}{l}{Scan Dir jackknife} \\ 
EE & 0.439 & 0.513 & 0.760 & 0.423 \\ 
BB & 0.944 & 0.535 & 0.565 & 0.168 \\ 
EB & 0.539 & 0.192 & 0.912 & 0.980 \\ 
\multicolumn{5}{l}{Tag Split jackknife} \\ 
EE & 0.543 & 0.537 & 0.810 & 0.938 \\ 
BB & 0.768 & 0.780 & 0.687 & 0.539 \\ 
EB & 0.313 & 0.547 & 0.407 & 0.451 \\ 
\multicolumn{5}{l}{Tile jackknife} \\ 
EE & 0.234 & 0.477 & 0.395 & 0.709 \\ 
BB & 0.050 & 0.072 & 0.012 & 0.046 \\ 
EB & 0.828 & 0.902 & 0.812 & 0.822 \\ 
\multicolumn{5}{l}{Phase jackknife} \\ 
EE & 0.862 & 0.982 & 0.577 & 0.471 \\ 
BB & 0.944 & 0.521 & 0.639 & 0.325 \\ 
EB & 0.691 & 0.890 & 0.204 & 0.357 \\ 
\multicolumn{5}{l}{Mux Col jackknife} \\ 
EE & 0.084 & 0.146 & 0.182 & 0.337 \\ 
BB & 0.172 & 0.337 & 0.012 & 0.152 \\ 
EB & 0.541 & 0.695 & 0.956 & 0.812 \\ 
\multicolumn{5}{l}{Alt Deck jackknife} \\ 
EE & 0.098 & 0.076 & 0.030 & 0.036 \\ 
BB & 0.092 & 0.126 & 0.102 & 0.140 \\ 
EB & 0.858 & 0.842 & 0.858 & 0.741 \\ 
\multicolumn{5}{l}{Mux Row jackknife} \\ 
EE & 0.232 & 0.289 & 0.699 & 0.918 \\ 
BB & 0.289 & 0.267 & 0.082 & 0.014 \\ 
EB & 0.148 & 0.130 & 0.996 & 0.998 \\ 
\multicolumn{5}{l}{Tile/Deck jackknife} \\ 
EE & 0.924 & 0.956 & 0.162 & 0.399 \\ 
BB & 0.507 & 0.034 & 0.561 & 0.343 \\ 
EB & 0.477 & 0.361 & 0.954 & 0.994 \\ 
\multicolumn{5}{l}{Focal Plane inner/outer jackknife} \\ 
EE & 0.477 & 0.335 & 0.200 & 0.792 \\ 
BB & 0.886 & 0.437 & 0.762 & 0.569 \\ 
EB & 0.595 & 0.876 & 0.926 & 0.780 \\ 
\multicolumn{5}{l}{Tile top/bottom jackknife} \\ 
EE & 0.261 & 0.519 & 0.998 & 0.990 \\ 
BB & 0.756 & 0.890 & 0.415 & 0.431 \\ 
EB & 0.850 & 0.920 & 0.377 & 0.317 \\ 
\multicolumn{5}{l}{Tile inner/outer jackknife} \\ 
EE & 0.184 & 0.353 & 0.427 & 0.529 \\ 
BB & 0.772 & 0.772 & 0.749 & 0.707 \\ 
EB & 0.407 & 0.038 & 0.934 & 0.667 \\ 
\multicolumn{5}{l}{Moon jackknife} \\ 
EE & 0.569 & 0.701 & 0.228 & 0.251 \\ 
BB & 0.305 & 0.465 & 0.978 & 0.990 \\ 
EB & 0.349 & 0.507 & 0.677 & 0.301 \\ 
\multicolumn{5}{l}{A/B offset best/worst} \\ 
EE & 0.635 & 0.267 & 0.104 & 0.431 \\ 
BB & 0.407 & 0.387 & 0.677 & 0.287 \\ 
EB & 0.321 & 0.605 & 0.860 & 0.685 \\ 

\end{tabular}
\end{ruledtabular}
\end{table}
\endgroup

\section{150\,GHz Spectral Stability}
\label{app:150}

\begin{figure}[htb]
\begin{center}
\resizebox{\columnwidth}{!}{\includegraphics{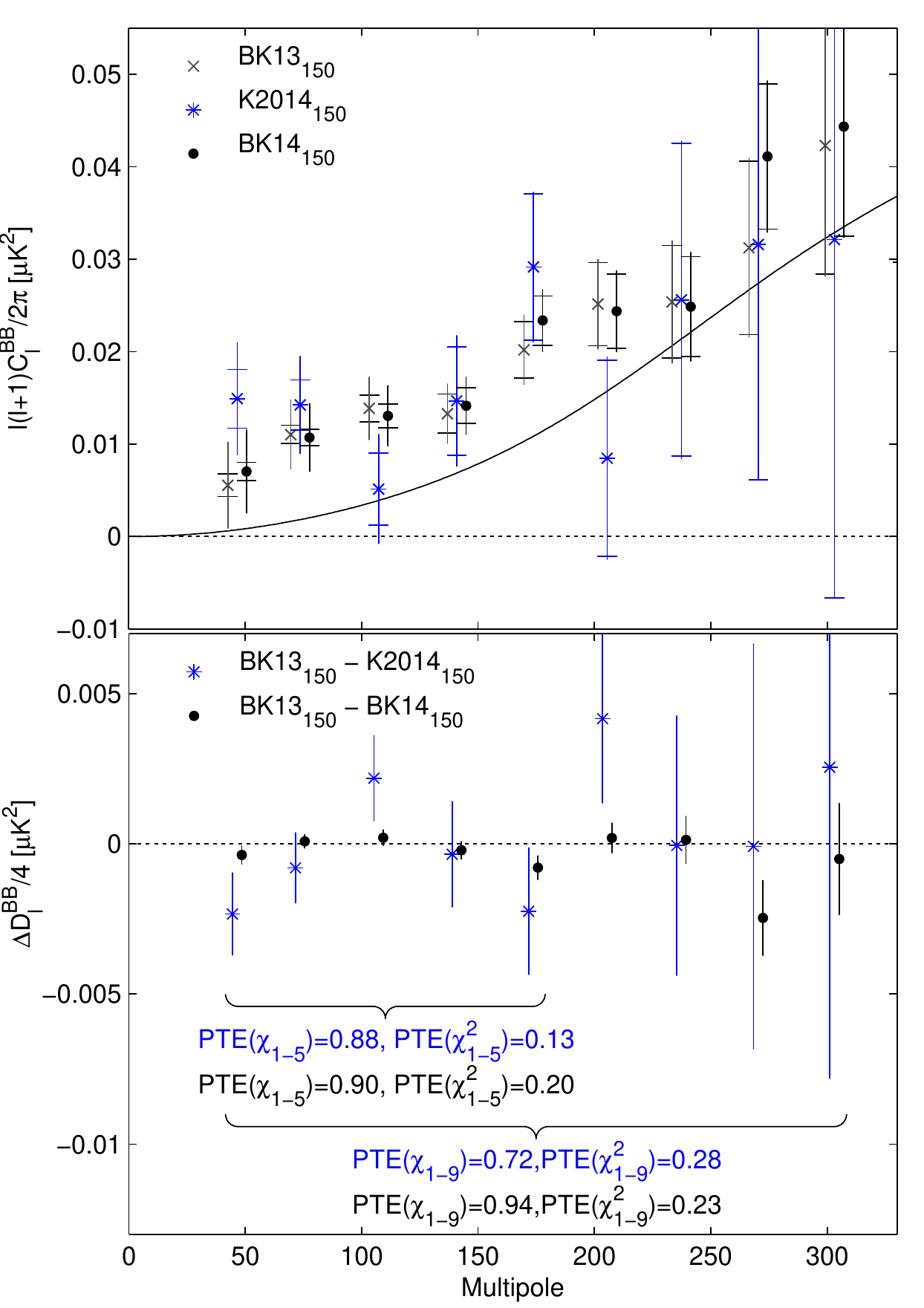}}
\end{center}
\caption{{\it Upper:} Comparison of the 150\,GHz $BB$ auto-spectrum
as previously published (BK13$_{150}$), for 2014 alone (K2014$_{150}$)
and for the addition of the two (BK14$_{150}$).
The inner error bars are the standard deviation of the lensed-\lcdm+noise 
simulations, while the outer error bars include the additional
fluctuation induced by a signal 
contribution matching the excess above lensing seen in the data.
Note that neither of these uncertainties are appropriate for comparison
of the band power values---for this see the lower panel.
(For clarity both sets of points are offset horizontally.)
{\it Lower:} The difference of the pairs of spectra shown 
in the upper panel divided by a factor of four.
The error bars are the standard deviation of the pairwise
differences of signal+noise simulations
which share common input skies (the simulations used to
derive the outer error bars in the upper panel).
Comparison of these points with null is an appropriate test
of the compatibility of the spectra and the PTE of $\chi$
and $\chi^2$ are shown.
This figure is similar to Fig.~8 of BK-V.}
\label{fig:specjack_150}
\end{figure}

Questions were raised as to whether the \biceptwo\
and \keckarray\ 2012+2013 $BB$ spectra are mutually
compatible.
We investigated this in Sec.~8 of BK-V and concluded
that they are.
Here we perform a similar test on the difference
of the BK13$_{150}$ and BK14$_{150}$ spectra---i.e.\ 
when adding the additional 150\,GHz data from 2014.
We compare the differences of the real spectra
to the differences of simulations which share the same underlying
input skies.
Fig.~\ref{fig:specjack_150} shows the results.
While the bandpowers do shift around even when adding only $\sim20$\%
of additional data these shifts are seen to be consistent
with noise fluctuation.
We go on to perform one more test---we instead take the difference
of the BK13$_{150}$ and the 2014 only 150\,GHz spectrum (which
we refer to as K2014$_{150}$).
Since we are measuring the bandpower differences in units of the
expected shift given the degree of common data we expect, and
find, similar results.

\section{Additional Spectra}
\label{app:allspec}

\begin{figure*}
\begin{center}
\resizebox{\textwidth}{!}{\includegraphics{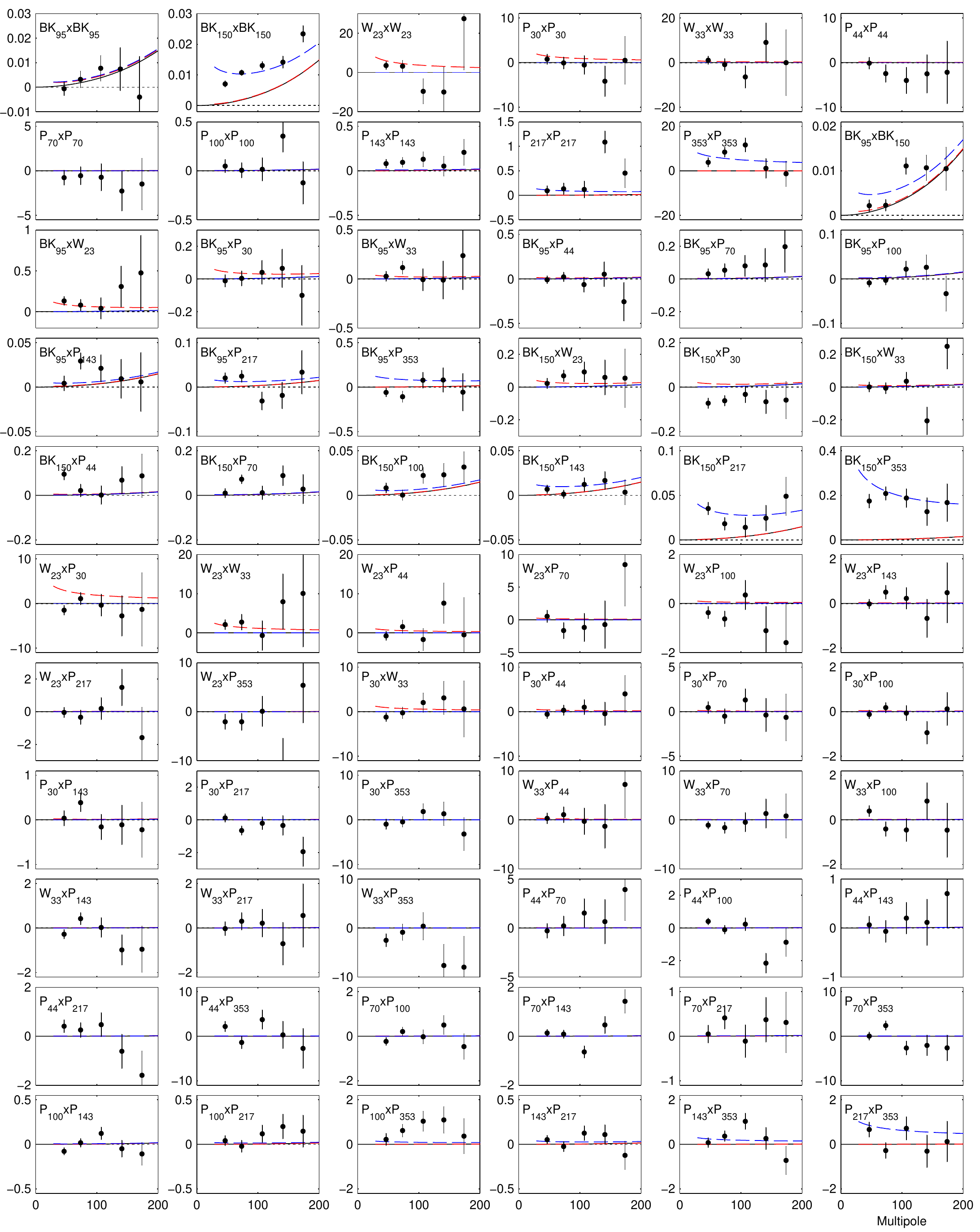}}
\end{center}
\caption{$BB$ auto- and cross-spectra between the
BK14 95 \&~150\,GHz maps and bands of \wmap\ and \planck.
In all cases the quantity plotted is $\clstar$ ($\mu$K$^2$),
and the black curves show the lensed-\lcdm\ theory spectrum.
The error bars are the standard deviations of the
lensed-\lcdm+noise simulations and hence contain no sample variance
on any additional signal component.
The blue dashed lines show a baseline lensed-\lcdm+dust model
($\Adf=4.3$\,$\mu$K$^2$, $\Bd=1.6$,
$\ad=-0.4$).
The red dashed lines show an upper limit lensed-\lcdm+synchrotron model
($\Asf=3.8$\,$\mu$K$^2$, $\Bs=-3.1$,
$\as=-0.6$).}
\label{fig:powspec_all}
\end{figure*}

Figures~2 \&~3
show only a small subset of the spectra which are used in the
likelihood analysis and included in the \texttt{COSMOMC}
input file.
We are using two \biceptwo/\keck\ bands, two \wmap\ bands,
and seven \planck\ bands resulting in 11 auto and 55 
cross-spectra.
In Fig.~\ref{fig:powspec_all} we show all of these
together with the baseline lensed-\lcdm+dust and
upper limit lensed-\lcdm+synchrotron models.
Note that, as expected from Fig.~8, several
spectra contribute to constraining synchrotron.

\begin{figure}
\begin{center}
\resizebox{0.7\columnwidth}{!}{\includegraphics{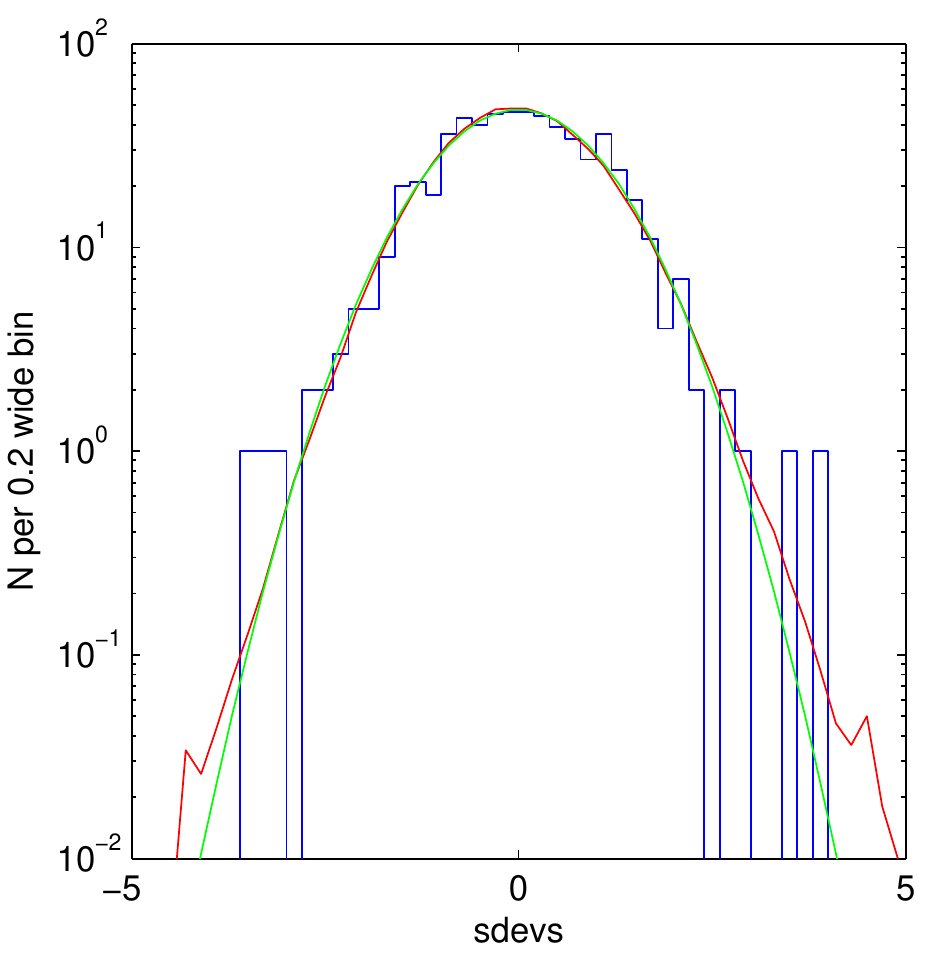}}
\end{center}
\caption{The normalized deviations of the bandpowers
shown in Fig.~\ref{fig:powspec_all} from the maximum
likelihood model is shown as the blue histogram.
The red curve is the same thing accumulated over 499 sims
of a lensed-\lcdm+dust model,
and the green curve shows a Gaussian with unit width.}
\label{fig:devs_all_hist}
\end{figure}

Fig.~\ref{fig:devs_all_hist} shows the distribution
of the normalized deviations between the data and the
maximum likelihood (ML) model (i.e. data minus expectation value
divided by the square root of the diagonal of the bandpower covariance
matrix).
Since the bandpower distributions are not strictly
Gaussian we overplot the same quantity from a
set of lensed-\lcdm+dust+noise simulations evaluated
against their input model.
(These simulations use the model $\Adf=3.75$\,$\mu$K$^2$,
$\beta_d=1.59$ and $\alpha_d=-0.42$.)
We see one nominally $4.0 \sigma$ point which is
bandpower four of P$_{217}\times$P$_{217}$ (see
Fig.~\ref{fig:powspec_all})---comparing to the simulated
distribution this event it not unlikely.
Taking $\chi^2$ versus the ML model yields
654, which compared to the distribution from simulations
has a PTE of $\sim0.1$.
We conclude that there is no evidence that the
signal or noise models are an inadequate explanation
of the data.

\section{Likelihood Variation and Validation}

\subsection{Likelihood Evolution}
\label{app:likeevol}

In Fig.~5 some evolutionary steps
were shown between the previous BKP analysis and the
new BK14 analysis presented in this paper.
Fig~\ref{fig:likeevol} shows some additional detail.
The first step is to the alternate analysis including synchrotron
which was shown in Fig.~8 of BKP (solid red to dashed-red).
This used the BK13 maps plus all of the polarized bands of \planck\
and set $\Bs=-3.3$ and $\as=-0.6$.
(In BKP the synchrotron pivot frequency was set to 150\,GHz
but since a fixed value of $\Bs$ was used there we can simply
transform the results to the pivot of 23\,GHz used in this work.)
Next we show the cumulative effects of model changes which
we have made for this paper:

\begin{figure*}[htb]
\begin{center}
\resizebox{1.0\textwidth}{!}{\includegraphics{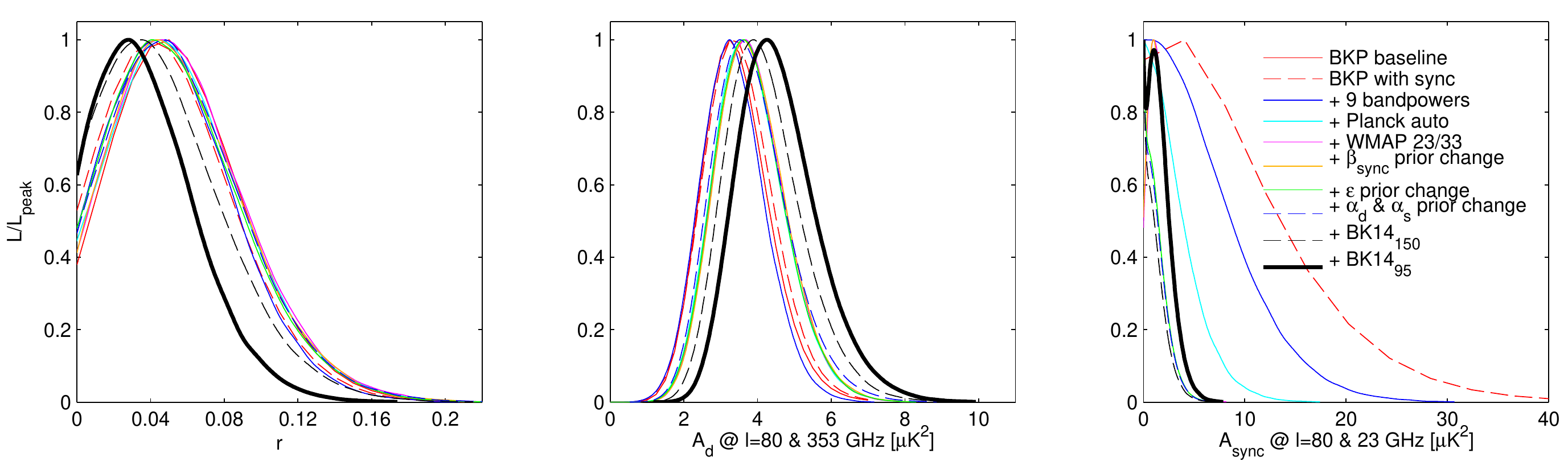}}
\end{center}
\caption{Evolution of the BKP analysis to the ``baseline'' analysis as
defined in this paper---see Appendix~\ref{app:likeevol} for details.}
\label{fig:likeevol}
\end{figure*}

We extend the bandpower range
from five ($20<\ell<200$) to nine ($20<\ell<330$) bandpowers---given that
lensing is included in the model there is no real reason not
to include these additional bandpowers
(dashed-red to solid blue).
We see that the $\As$ constraint tightens somewhat.

We switch from the use of \planck\ single-frequency split/split
cross-spectra (in this case Y1$\times$Y2) to full map auto spectra
(blue to cyan).
This is done for technical reasons---substituting in the
cross-spectra causes numerical problems in the HL likelihood.
The auto spectra have higher signal-to-noise and the
constraint on $\As$ tightens further.

We include the \wmap\ 23 \& 33\,GHz bands and see that these
have considerable additional power to constrain synchrotron
(cyan to magenta).

In BKP we used $\Bs=-3.3$ as this is the mean value within
our field of the ``model f'' synchrotron spectral
index maps available for download from the \wmap\
website~\footnote{See \url{http://lambda.gsfc.nasa.gov/product/map/dr5/mcmc_maps_info.cfm}}.
However that analysis does not distinguish between the
spectral behavior of temperature and polarization
anisotropy.
Ref.~\cite{fuskeland14} analyzed the \wmap\ data and
found a mean value of $\Bs=-3.1\pm0.04$ for polarization at high galactic
latitude.
In this analysis we use a central value of $\Bs=-3.1$,
and since possible patch-to-patch variation is poorly constrained,
to be conservative we marginalize over a Gaussian prior with
width $\sigma=0.3$.
More recently Ref.~\cite{choi15} examined the same
data and found $\Bs \approx -3.0$ with considerable fluctuation.
This change has very little effect
(magenta to yellow).

Polarized synchrotron and dust emission can be
spatially correlated---indeed they are guaranteed
to be so on the largest scales.
Ref.~\cite{choi15} reports a correlation of 0.2
for $30<\ell<200$.
To be conservative in this analysis we marginalize
over the range $0<\epsilon<1$.
This causes the constraint
on synchrotron to tighten because of the non-detection of
signal in spectra like P$_{30}\times$P$_{353}$
(yellow to green).
We note that the data prefer the value $\epsilon=0$
as seen in the upper-right panel of Fig.~4.

In BKP we used $\ad=-0.42$ following the analysis
of large regions of high latitude sky in Ref.~\cite{planckiXXX},
and $\as=-0.6$ taken from Ref.~\cite{dunkley08}.
In this work we found that we can marginalize
over generous ranges in these parameters
$-1<\ad<0$ \& $-1<\as<0$ with only a tiny
change in the bottom line results so we choose to do so
(green to dashed-blue).

Finally we show the changes resulting from adding the
new 150\,GHz and 95\,GHz data (dashed-blue to dashed-black
and dashed-black to heavy-black).
As already seen in Fig.~5 these are much more
significant.

\subsection{Likelihood Variation}
\label{app:likevar}

\begin{figure*}[htb]
\begin{center}
\resizebox{1.0\textwidth}{!}{\includegraphics{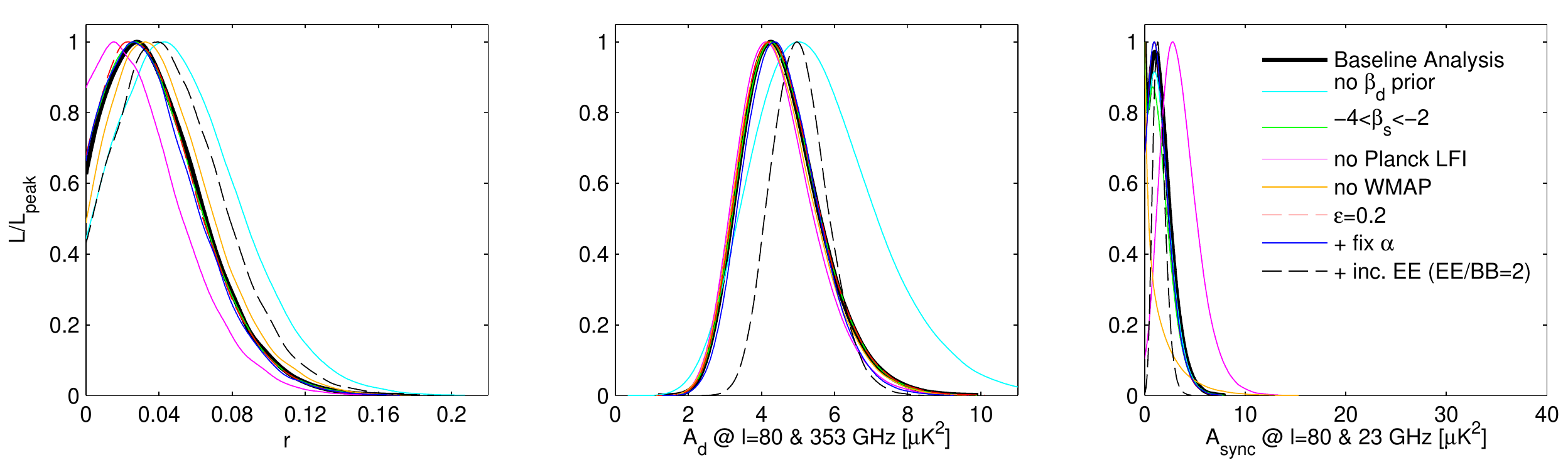}}
\end{center}
\caption{Likelihood results when varying the data sets and the
model priors---see Appendix~\ref{app:likevar} for details.}
\label{fig:likevar}
\end{figure*}

In Fig.~\ref{fig:likevar} we investigate several variations
to the baseline analysis in terms of the model priors and
input data sets.
The first four of these loosen the priors and/or remove data,
while the final three tighten the priors and/or add data.

First we repeat a variation already shown in
Fig.~5---we remove the prior on the
frequency spectral index of dust $\Bd$
(black to cyan).
The data then constrains $\Bd$ to a well behaved, approximately
Gaussian range (not shown) with mean/$\sigma$ of 1.82/0.26.
The value of $\Adf$ shifts up slightly but, with the steeper slope
versus frequency, the $r$ constraint also shifts up slightly to
$r=0.043^{+0.033}_{-0.031}$ with a zero-to-peak likelihood
ratio of 0.44 (10\% likely if $r=0$).

Second we relax the prior on the frequency
spectral index of synchrotron to $-4<\Bs<-2$ and see
that this has very little effect on any of the curves
(black to green).

Third we remove all the \planck\ LFI bands from consideration
(black to magenta).
This causes the peak of the $r$ constraint to shift down
a little and the $\As$ constraint to peak quite strongly away
from zero, while the $\Ad$ constraint is not significantly affected.

Fourth we drop the two bands of \wmap\
(black to yellow).
This slightly decreases the zero-to-peak ratio
of the $r$ constraint and significantly tightens
the $\As$ constraint.

We now progressively tighten the priors.
For the fifth curve we switch from $0<\epsilon<1$ to the
value preferred by Ref.~\cite{choi15} $\epsilon=0.2$
(black to dashed-red).
This makes almost no difference to any of the constraints,
although we do note that the up-tick in the $\As$ curve
approaching zero goes away.

In the sixth curve we also go back to the tight priors
$\ad=-0.42$ and $\as=-0.6$ which were used in
the BKP analysis (dashed-red to blue).
As expected from Fig.~\ref{fig:likeevol} this makes
almost no difference to any of the constraints.

Finally in the seventh curve we also include all the $EE$ and $EB$ spectra
under the assumption that the $EE/BB$ ratios for
dust and synchrotron are exactly 2
(blue to dashed-black).
For dust this ratio was found to apply when
averaging over large areas of sky in Ref.~\cite{planckiXXX}.
Ref.~\cite{choi15} states that this ratio also
applies on average for synchrotron.
Assuming this fixed ratio leads to extra constraining
power---the $r$ curve shifts up slightly, the $\Ad$
curve narrows and the $\As$ curve peaks strongly away from zero.
It is unclear how much patch-to-patch variation
we should in fact allow in the $EE/BB$ ratio so this
variation should not be over interpreted at this time.

\subsection{Likelihood Validation}
\label{app:likevalid}

\begin{figure*}[htb]
\begin{center}
\resizebox{0.8\textwidth}{!}{\includegraphics{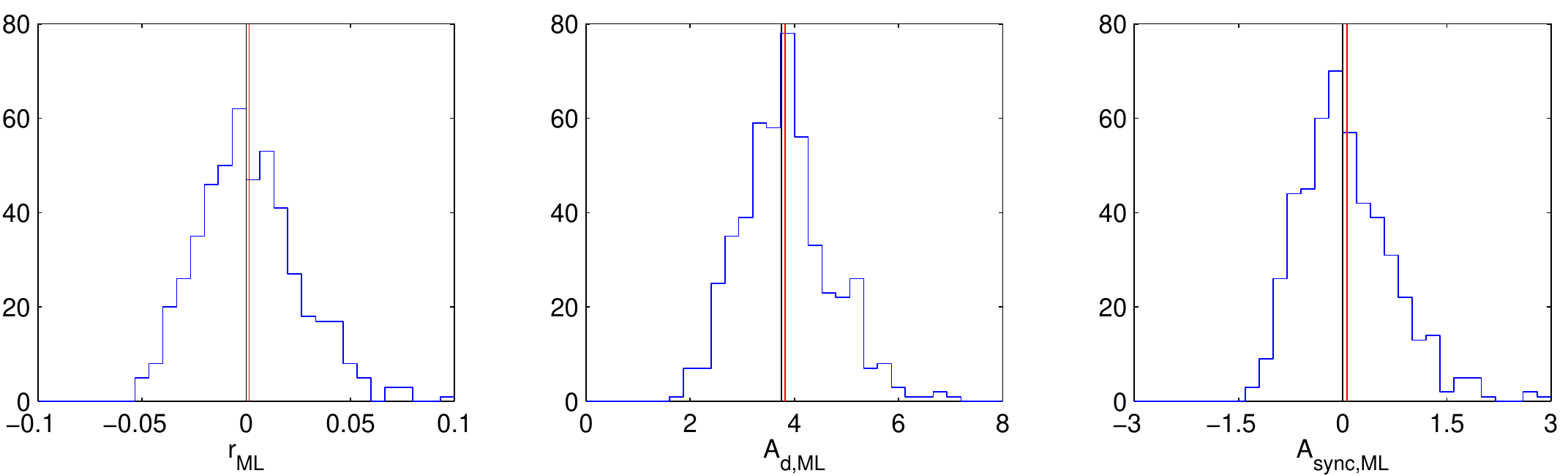}}
\end{center}
\caption{Results of validation tests running the
likelihood on simulations of a lensed-\lcdm+dust model
($A_{d,353}=3.75$\,$\mu$K$^2$, $\beta_d=1.59$ and $\alpha_d=-0.42$).
The blue histograms are the recovered ML values with the
red line marking their means.
The black line shows the input value.
In the left panel $\sigma(r)=0.024$.
See Appendix~\ref{app:likevalid} for details.}
\label{fig:likevalid}
\end{figure*}

As already mentioned we run full 
timestream simulations
of a lensed-\lcdm+dust model ($A_{d,353}=3.75$\,$\mu$K$^2$,
$\beta_d=1.59$ and $\alpha_d=-0.42$).
We would like to check that the HL likelihood as implemented
is capable of recovering the input values of this model.
However if we run the standard COSMOMC analysis on these
we of course find that the ML values are biased, since only
zero or positive values of $r$ and $\As$ are allowed.
We therefore instead run a ML search on each sim realization
where the values of $r$ and $\As$ are artificially allowed to
go negative (as is $\Ad$ although in practice it doesn't).
Fig.~\ref{fig:likevalid} shows the results---the input values
are recovered in the mean as expected.

An additional piece of information which comes from
this study is the standard deviation of the 
recovered ML $r$ parameter, $\sigma(r)=0.024$.
Unlike the width of the 68\% highest posterior density intervals
derived from the marginalized $r$ curve shown in Fig.~4
and quoted with our baseline results,
this $\sigma(r)$ statistic is insensitive to where the peak value
preferred by the data happens to lie, and is therefore
a more robust measure of the intrinsic constraining
power of the experimental data.

\end{appendix}

\end{document}